\documentclass[%
nofootinbib,
 amsmath,amssymb,
 aps,
]{aa}

\usepackage[varg]{txfonts}
\usepackage{graphicx}
\usepackage{dcolumn}
\usepackage{bm}
\usepackage[hidelinks]{hyperref}
\usepackage[mathlines]{lineno}
\usepackage[english]{babel}
\usepackage{amssymb}
\usepackage{amsmath}
\usepackage{latexsym}
\usepackage{cancel}
\usepackage{xcolor}
\usepackage{esint}
\usepackage{caption}
\usepackage{subcaption}

\usepackage[normalem]{ulem}

\bibpunct{(}{)}{;}{a}{}{,}

\begin{document}

\title{Covariant formulation of refracted gravity}

\author{Andrea P. Sanna\inst{\ref{inst1},\ref{inst2},\ref{inst3}} \and Titos Matsakos\inst{\ref{inst3}} \and Antonaldo Diaferio\inst{\ref{inst3},\ref{inst4}}}



\institute{
Dipartimento di Fisica, Universit\`a di Cagliari, Cittadella Universitaria, 09042, Monserrato, Italy \email{asanna@dsf.unica.it}\label{inst1}
\and
Istituto Nazionale di Fisica Nucleare (INFN), Sezione di Cagliari, Cittadella Universitaria, 09042, Monserrato, Italy\label{inst2}
\and 
Dipartimento di Fisica, Universit\`a di Torino, via P. Giuria 1, 10125, Torino, Italy \email{titos.matsakos@gmail.com, antonaldo.diaferio@unito.it}\label{inst3}
\and
Istituto Nazionale di Fisica Nucleare (INFN), Sezione di Torino, Torino, Italy\label{inst4}
} 

\date{Received / Accepted}

\abstract
{

We propose a covariant formulation of refracted gravity (RG), which is a classical theory of gravity based on the introduction of  gravitational permittivity --- a monotonic function of  the local mass density --- in the standard Poisson equation. Gravitational permittivity mimics  dark matter phenomenology. The covariant formulation of RG (CRG) that we propose belongs to the class of scalar-tensor theories, where the scalar field $\varphi$ has a self-interaction potential $\mathcal{V}(\varphi)=-\Xi\varphi$, with $\Xi$ being a normalization constant. We show that the scalar field is twice the gravitational permittivity in the weak-field limit. Far from a spherical source of density $\rho_{\mathrm s}(r)$, the transition between the Newtonian and the RG regime appears below the acceleration scale $a_\Xi=(2\Xi-8\pi G\rho/\varphi)^{1/2}$, with $\rho=\rho_\mathrm{s}+\rho_\mathrm{bg}$ and $\rho_\mathrm{bg}$ being an isotropic and homogeneous background. In the limit $2\Xi\gg 8\pi G\rho/\varphi$,  we obtain $a_\Xi\sim 10^{-10}$~m~s$^{-2}$. This acceleration is comparable to the acceleration $a_0$ originally introduced in MOdified Newtonian Dynamics (MOND). 
From CRG, we also derived the modified Friedmann equations  for an expanding, homogeneous, and isotropic universe. We find that the same scalar field $\varphi$ that mimics dark matter also drives the accelerated expansion of the Universe. From the stress-energy tensor of $\varphi$, we derived the equation of state of a redshift-dependent effective dark energy $w_\mathrm{DE}=p_\mathrm{DE}/\rho_\mathrm{DE}$. Current observational constraints on $w_\mathrm{DE}$ and distance modulus data of type Ia supernovae suggest that $\Xi$ has a comparable value to the cosmological constant $\Lambda$ in the standard model. Since $\Xi$ also plays the same role of $\Lambda$, CRG suggests a natural explanation of the known relation $a_0\sim \Lambda^{1/2}$. 
CRG thus appears to describe both the dynamics of cosmic structure and the expanding Universe with a single scalar field, and it falls within the family of models that unify the two dark sectors, highlighting a possible deep connection between phenomena currently attributed to dark matter and dark energy separately. 
}


\maketitle

\section{Introduction}

The current standard  $\Lambda$ cold dark matter ($\Lambda$CDM) cosmological model assumes that gravitational interactions are ruled by general relativity (GR); the model relies on the existence of collisionless non-baryonic CDM, and a positive cosmological constant $\Lambda$ \citep{1995Natur.377..600O}. Non-baryonic dark matter is required to account for the abundance of light elements \citep{2016RvMP...88a5004C}, the amplitude of the power spectrum of the temperature anisotropies in the cosmic microwave background (CMB) \citep{Aghanim:2019ame}, and the dynamics of cosmic structures \citep{Clowe:2006eq, Dodelson:2001ux, DelPopolo:2013qba, Akrami:2018mcd, Planck:2018nkj}. The cosmological density parameter $\Omega_{\Lambda0}=\Lambda/3H_0^2$ associated with $\Lambda$, with $H_0$ being the Hubble constant, accounts for the negative value of the deceleration parameter $q_0$ measured with the Hubble diagram of type Ia supernovae (SNeIa) \citep{Riess+98, Perlmutter+99}. The curvature of the Universe, $\Omega_k=\Omega_{\Lambda0}+\Omega_0-1$, measured from the CMB power spectrum, suggests a null curvature \citep{Aghanim:2018eyx}.
 
Although the $\Lambda$CDM model agrees with most of the rich observational information currently available,  a number of tensions are present both on large cosmic scales and on the scale of galaxies. The Hubble constant $H_0$ estimated with the distance ladder in the local Universe \citep{2016ApJ...826...56R, 2019ApJ...876...85R} is more than $4\sigma$ larger than $H_0$ inferred from the measurements of the CMB \citep{Verde:2019ivm}. The observed abundances of the light elements are consistent with the Big Bang nucleosynthesis scenario, except $^7$Li, whose abundance is a factor of $\sim 3$ smaller \citep{Mathews:2019hbi}. Numerous features of the CMB temperature anistropies are present on large scales \citep{Ade:2015hxq}. The probability of some of these features to appear individually is $\lesssim 0.1\%$; the combined probability of the uncorrelated features is $\lesssim 0.001\%$ and might represent a serious challenge to the $\Lambda$CDM model \citep{Schwarz:2015cma, Luongo:2021nqh}. The lensing amplitude in the CMB power spectra is enhanced  compared to $\Lambda$CDM expectations \citep{Aghanim:2019ame} and would suggest a positive rather than a null curvature of the Universe \citep{Handley:2019tkm, DiValentino:2019qzk}. 
A slight tension also appears for the normalization of the power spectrum $\sigma_8$ \citep[e.g.][]{Raveri:2015maa}: the value inferred from the CMB measurements \citep{Aghanim:2018eyx} is $2\sigma$ larger than the value derived from the tomographic weak gravitational lensing analysis of the Kilo-Degree Survey (KiDS) imaging data \citep{Hildebrandt:2016iqg}.

In addition, the cosmological constant poses a fine-tuning problem that is theoretically challenging. 
 If we associate $\Lambda$ with the ground state energy level of the vacuum in quantum field theory, its measured value, $\Lambda\sim 10^{-12}$~eV$^4$, appears  to be $\sim 120$ orders of magnitude smaller than expected \citep{Weinberg+89, Padilla:2015aaa}. If we associate $\Lambda$ with the energy scale up to which the standard model of particle physics has been tested, $\sim 1$~TeV, the discrepancy reduces to $\sim 60$ orders of magnitude \citep{Joyce+15}, but it remains severe. The most popular solution to the $\Lambda$ problem is to suppress $\Lambda$ in the Einstein-Hilbert action and generate the accelerated expansion of the Universe with dark energy, an auxiliary scalar field with proper features. The specific implementation of this idea has generated a large number of different models that may or may not modify Einstein's equations \citep[see][for extensive reviews]{Peebles+03, Copeland:2006wr, Bamba:2012cp, Joyce+15, Amendola:2016saw}.

Collisionless CDM poses additional problems on small scales: the core-cusp problem in dwarf and disk galaxies, the missing satellite problem, the too-big-to-fail problem, and the plane of satellite galaxies in the Milky Way and nearby large galaxies \citep{Salucci:2003sa, 2012MNRAS.425.2817F, Boylan-Kolchin:2011lmk, Garrison-Kimmel:2014vqa, DelPopolo+17, Kroupa2012, deMartino:2020gfi}. In addition, some relations, such as the radial acceleration relation or the baryonic Tully-Fisher relation in disk galaxies, would require finely tuned interactions between CDM and baryonic matter \citep{Desmond:2015nja, DiCintio:2015eeq, Desmond:2016azy}.

Some of these small-scale tensions may originate either by an inaccurate treatment of the dynamics of CDM and baryonic matter or by the inappropriate properties adopted for the dark matter model.  For example, the core-cusp problem emerges when we attempt to interpret the observed kinematics of stars in galaxies by assuming that the galaxies are embedded within CDM halos with a Navarro-Frenk-White (NFW) density profile,  as predicted by N-body simulations \citep{Navarro:1996gj}. By dropping this constraint on the dark matter distribution and adopting a dark matter density profile with a flat central core, we can properly describe the stellar kinematics of spirals with a universal rotation curve \citep{Persic:1995ru, Salucci:2007tm, Salucci:2012zcw, Salucci:2018eie}. Indeed, an exponential disk and a dark matter halo described by the Burkert profile with a core \citep{Burkert:1995yz} excellently describe the rotation curves of five spirals \citep{Gentile:2004tb}, and suggest that spirals and dwarf galaxies lie on the same scaling relation between the core density and the core radius of the dark matter halo \citep{Salucci:2000ps}. A number of different effects from the dynamics of CDM or from baryonic physics, including stellar feedback and star formation efficiency,  are  advocated to generate a central core in the dark matter distribution \citep{deMartino:2020gfi}. For example, tidal effects by a massive hosting galaxy might induce dark matter density profiles shallower than the NFW profile in the central regions of satellite halos and might also alleviate the too-big-to-fail problem \citep{Tomozeiu2016}.

Dropping the hypothesis that dark matter is collisionless and cold might solve some, albeit not all, of the tensions of CDM on the scale of galaxies \citep{Salucci:2018hqu, DiPaolo:2020tem}. Indeed, weakly interacting massive particles (WIMPs) are the most plausible candidate to make up the collisionless CDM \citep{Bertone+05}. However, attempts to, directly or indirectly, detect these particles have not yet produced unquestionable results \citep{Tanabashi:2018oca}. Alternative dark matter models include warm dark matter, self-interacting dark matter, QCD axions, and fuzzy dark matter  \citep{deMartino:2020gfi}. Some of these models have particle counterparts,  such as sterile neutrinos or ultra-light bosons. The detection of these particles requires direct or indirect experiments different from those conceived for detecting WIMPs \citep{Buckley:2017ijx}. Dark matter particles with distinct peculiar features, such as superfluidity \citep{Berezhiani:2015bqa} or gravitational polarization \citep{Blanchet:2006yt, Blanchet:2017duj}, can also partly reproduce the phenomenology of galaxies.

Alternatively, the dynamics on the scale of galaxies could be explained by modifying the theory of gravity in the
weak-field Newtonian limit without resorting to the existence of dark matter. The idea of MOdified Newtonian Dynamics (MOND) was originally motivated only by the observations of the flat rotation curves of disk galaxies \citep{Milgrom1, Begeman:1991iy, Begum:2003eu}, and  many of the current observational challenges on the scale of galaxies were actually predicted by MOND \citep{Milgrom2,Milgrom3, Sanders:2002pf}. The universal rotation curves of spirals \citep{Persic:1995ru} and the universality of the galactic surface density within the radius of the core of a Burkert dark matter profile also appear to be consistent with MOND \citep{Gentile:2008rv, Gentile:2009bw}, although the debate on the dynamics of dwarf disk galaxies remains vibrant \citep{2007MNRAS.374.1051C, Sanchez-Salcedo:2013gql, DiPaolo:2018mae, Banik:2020kwt, DiPaolo:2020tem}. Furthermore, the MOND formulation is purely phenomenological and its extension to a covariant formulation has proven to be challenging \citep{Famaey+12, MilgromReview, Skordis:2019fxt, Zlosnik:2017xpr, Skordis:2020eui}. 

Although the problems of the cosmological constant and dark matter are usually considered two separate issues, attempts to unify the two dark sectors in a single framework are numerous. 
For example, \citet{Ferreira:2018wup} suggest a model where the dark matter is made of two superfluids arising from two distinguishable states separated by a small energy difference: particles of one species can be converted into the other and the interaction between these multi-state components of the dark matter can drive cosmic acceleration.
Alternatively, in emergent gravity,  where  both the classical space-time structure and gravity emerge from an underlying microscopic quantum theory \citep{Sakharov:1967pk, Padmanabhan:2014jta, Verlinde:2016toy}, the two dark sectors can be unified when, for example, the dark energy fluid is modelled as a critical Bose-Einstein condensate of gravitons \citep{Cadoni:2017evg, Cadoni:2018dnd, Tuveri:2019zor, Cadoni:2020izk}.

Attempts to describe both the accelerated expansion of the Universe and the dynamics of cosmic structures without dark matter  and dark energy by building a modified theory of gravity are also numerous \citep[see, e.g.,][]{Clifton+12, Nojiri:2017ncd}. According to Lovelock's theorem \citep{Lovelock:1971yv, Lovelock:1972vz}, we can build a metric theory of gravity different from GR by, for example, allowing derivatives of the metric tensor higher than second order in the field equations, or introducing other fields in addition to the metric tensor. 

Conformal gravity adopts the former route and replaces the Einstein-Hilbert action with the contraction of the fourth-rank conformal tensor introduced by Weyl \citep{Mannheim:1992tr}. Conformal gravity does not present ghost-like instabilities, that might be common in theories with high-order derivatives \citep{Bender:2007wu}, and is renormalizable \citep{Mannheim:2011ds}. Unfortunately, although conformal gravity successfully reproduces the accelerated expansion of the Universe \citep{Diaferio:2011kc}, the expected abundance of primordial deuterium is orders of magnitudes smaller than observed \citep{Elizondo:1994vh}. Moreover, conformal gravity is unable to reproduce the dynamics of galaxy clusters \citep{8219260, Diaferio:2008gh},  and appears to require a fine-tuning condition to describe the phenomenology of gravitational lensing and the dynamics of disk galaxies \citep{2019CQGra..36x5014C}.

We can preserve the second-order field equations by introducing a single scalar field that drives both the accelerated expansion of the Universe and the formation of cosmic structure \citep[e.g.][]{Carneiro:2018zet}. The case of a classical scalar field with a non-canonical kinetic term in its Lagrangian generates the class of Unified Dark Matter models \citep{Bertacca:2010ct}. These models can be a viable alternative to $\Lambda$CDM \citep{Camera:2010wm, Camera:2017tws} if the effective sound speed is small enough that the scalar field can cluster \citep{Bertacca:2008uf, Camera:2009uz}.

Here, we contribute to the quest for a unified model of the two dark sectors by proposing a novel scalar-tensor theory \citep{Fujiibook, Quiros2019} where the scalar field is responsible for both the dynamics of cosmic structures and the accelerated expansion of the Universe.
This scalar-tensor theory is the covariant formulation of refracted gravity (RG), which is a new phenomenological modified theory of gravity proposed by \citet{MatsakosDiaferio16}. Refracted gravity appears to reproduce the phenomenology on the scale of galaxies and galaxy clusters by introducing a monotonic function of the local mass-density in the standard Poisson equation, termed gravitational permittivity. Indeed, \citet{Cesare+20} showed that RG properly describes the rotation curves and the vertical velocity dispersion profiles of 30 disk galaxies in the DiskMass Survey \citep{Bershady:2010sj}, and the dynamics of stars and globular clusters in the outer regions of three elliptical galaxies of type E0 \citep{Cesare+21}. Here, we provide a covariant formulation  of this non-relativistic formulation of RG.

Section ~\ref{Sec:NonRelativisticRG} reviews the relevant features of RG. In Sect.~\ref{Sec:RRGtheory}, we derive the covariant formulation of RG (CRG, hereafter) in the framework of scalar-tensor theories, and show that the scalar field can be identified with the gravitational permittivity. In Sect.~\ref{Sec:RRGCosmology}, we consider a homogeneous and isotropic universe and derive the modified Friedmann equations; we show that the scalar field is responsible for the accelerated expansion of the Universe, and derive the equation of state of a redshift-dependent effective dark energy. We conclude in Sect.~\ref{Sec:Conclusions}.

\section{Non-relativistic refracted gravity}
\label{Sec:NonRelativisticRG}

The phenomenological RG is based on the modified Poisson equation (hereafter RG equation) \citep{MatsakosDiaferio16}
\begin{equation}
\vec{\nabla} \cdot \left(\epsilon \vec{\nabla} \Phi \right) = 4 \pi G \rho\,,
\label{eq:NRRG}
\end{equation}
where $\Phi$ is the gravitational potential, $G$ the gravitational constant, and $\rho$ the density of ordinary matter. The function $\epsilon$ is the gravitational permittivity, according to the mathematical similarity with the term on the left-hand side of the Poisson equation that describes electric fields in matter. We emphasize that no other parallels have been drawn with electrodynamics. As a starting hypothesis, the permittivity was prescribed to be a monotonic function of the local mass density, that is $\epsilon(\rho)$. However, $\rho$ was only chosen because is the simplest scalar field characterizing the matter distribution in the weak-field regime; $\epsilon$ could in principle depend on other local quantities, for example the trace of the energy-momentum tensor or the entropy.

On scales where the visible mass can accurately explain the observed dynamics according to Newtonian gravity, for example on the scale of stars,  $\epsilon$ must be constant and equal to $1$ in order to recover the standard Newtonian Poisson equation, $\nabla^2 \Phi = 4\pi G \rho$. Therefore, we can adopt the form of the permittivity 
\begin{equation}
  \epsilon(\rho) \simeq \left\{
  \begin{array}{ll}
    1 & \mathrm{for}\ \ \rho \gtrsim \rho_\mathrm{thr} \\
    \epsilon_\mathrm{v} & \mathrm{for}\ \ \rho \ll \rho_\mathrm{thr}
  \end{array}
  \right.\,,
\label{eq:epsilon}
\end{equation}
where $\rho_\mathrm{thr}$ is the density threshold that sets the transition between the Newtonian regime and the modified gravity regime;\footnote{\citet{MatsakosDiaferio16} adopt the designation critical density $\rho_\mathrm{c}$ rather than threshold density $\rho_\mathrm{thr}$. Here, we prefer to adopt the latter, to avoid any confusion with the critical density of the Universe, that is used in modelling a homogeneous and isotropic universe. 
In addition, here $\rho_\mathrm{thr}$ is explicitly related to the value below which $\epsilon$ deviates from unity.}  $0<\epsilon_\mathrm{v}<1$ is the permittivity of the vacuum.

The presence of $\epsilon$ in the Poisson equation, Eq.~(\ref{eq:NRRG}), has two main effects: (1) in low-density regions, it generates a stronger gravitational field than in Newtonian gravity; and (2) it bends the gravitational field lines in regions where $\vec{\nabla}\Phi$ and $\vec{\nabla}\epsilon$ are not parallel. The former effect is trivially seen in the simple case $\epsilon=\mathrm{const}<1$, when $\rho < \rho_\mathrm{thr}$: Eq.~(\ref{eq:NRRG}) becomes $\nabla^2 \Phi = 4\pi G \rho/\epsilon$, and shows that the resulting field is equivalent to a Newtonian field originating from a larger effective mass density $\rho/\epsilon$, or a larger gravitational constant $G/\epsilon$. When $\vec{\nabla}\Phi$ and $\vec{\nabla}\epsilon$ are not parallel, the field lines are refracted; in other words, the field lines  change their direction when they cross the iso-surfaces of $\epsilon$ at an angle different from $\pi/2$. This effect can also generate non-Newtonian phenomena in regions where the density is larger than $\rho_\mathrm{thr}$, because the redirection of the field lines has non-local consequences. For example, when we consider only ordinary matter as the gravitational source,
RG predicts flat rotation curves in disk galaxies even in regions where $\rho > \rho_\mathrm{thr}$ \citep{Cesare+20}. The redirection of the field lines is also expected to be responsible for the mass discrepancy in galaxy clusters \citep{MatsakosDiaferio16}. In these two studies, $\epsilon(\rho)$ is assumed to be a smooth step function, with the density threshold $\rho_\mathrm{thr}$  in the  range $10^{-27}$--$10^{-24}$\,g\,cm$^{-3}$.

\section{Refracted gravity as a scalar-tensor theory}
\label{Sec:RRGtheory}

We present the fundamental equations of CRG in Sect. \ref{sec:FundEq}, and the relation of CRG with the Horndeski theory and the CRG screening mechanisms in Sect. \ref{sec:Horndeski} and Sect. \ref{sec:Screening}, respectively. In Sect. \ref{sec:weakfieldlimit}, we derive the weak field limit of CRG.

\subsection{Fundamental equations of CRG}
\label{sec:FundEq}

The family of scalar-tensor (TeS) theories, with the scalar field $\varphi$ non-minimally coupled to the rank-2 tensor field $g_{\mu\nu}$, can be derived from the action\footnote{Hereafter, we use natural units with $c = \hbar = 1$.}
\begin{align}
\mathcal{S} &= \frac{1}{16\pi G}\int d^4 x \ \sqrt{g} \left[\varphi R + \frac{\mathcal{W}(\varphi)}{\varphi}  \nabla^{\alpha}\varphi \nabla_{\alpha} \varphi + 2\mathcal{V}(\varphi)\right]+\nonumber\\
&\quad + \int d^4 x \sqrt{g} \ \mathcal{L}_\mathrm{m} (g_{\mu \nu}, \psi_\mathrm{m})\,,
\label{eq:TeSaction}
\end{align}
where $-g$ is the determinant of $g_{\mu\nu}$, $R = g^{\mu\nu}R_{\mu\nu}$ is the Ricci scalar, $R_{\mu\nu}$ is the Ricci tensor, $g^{\mu \nu}$ is the inverse metric, $\nabla_{\mu}$ is the covariant derivative, $\nabla^{\alpha} \varphi \nabla_{\alpha} \varphi \equiv g^{\alpha \beta} \nabla_{\alpha}\varphi \nabla_{\beta}\varphi$, and $\mathcal{L}_\mathrm{m}$ is the matter Lagrangian density, with $\psi_\mathrm{m}$ being the matter fields \citep{Faraoni+04}. The potential $\mathcal{V}(\varphi)$ and the general differentiable function of the scalar field $\mathcal{W}(\varphi)$ parametrize the family of TeS theories. When $\varphi\to 1$, the TeS action reduces to the standard Einstein-Hilbert action with a cosmological constant equal to $-\mathcal{V}(1)$. Hereafter, for the sake of simplicity, we consider the $\varphi$-dependence of $\mathcal{W}$ and $\mathcal{V}$ implicit.

Varying Eq.~(\ref{eq:TeSaction}) with respect to the metric yields the ten modified Einstein field equations in the Jordan frame (JF)\footnote{We adopt the sign conventions of \citet{weinberggravitation}, with the Ricci tensor given by $R_{\mu\nu} =\partial_{\nu} \Gamma^{\alpha}_{\mu \alpha} - \partial_{\alpha} \Gamma^{\alpha}_{\mu \nu} + \Gamma^{\alpha}_{\mu \beta} \Gamma^{\beta}_{\alpha \nu} - \Gamma^{\alpha}_{\mu \nu} \Gamma^{\beta}_{\alpha \beta}$, and the standard Einstein equations $G_{\mu\nu} = -8\pi G T_{\mu\nu}$.}$^,$ \footnote{
We choose to work in the JF rather than in the Einstein frame (EF) because, in the JF, the scalar field is interpreted as a modification to the gravitational field (the left-hand side of the field equations) rather than a modification to the stress-energy tensor as in the EF (the right-hand side of the field equations). In the weak-field limit, this modification is consistent with the Poisson equation of the non-relativistic formulation of RG (Eq.~\ref{eq:NRRG}), which contains modifications to the field, and not to the source term. As we show in Sect.~\ref{sec:weakfieldlimit}, this formulation naturally leads to the identification of the scalar field with the permittivity. 
}
\begin{align}
&\varphi \left(R_{\mu \nu} - \frac{1}{2} g_{\mu \nu} R \right)  - \left(\frac{\mathcal{W}}{2 \varphi} \nabla^{\alpha} \varphi \nabla_{\alpha} \varphi + \Box \varphi + \mathcal{V}\right)g_{\mu \nu}+ \nonumber\\
&\quad + \nabla_{\mu} \nabla_{\nu} \varphi + \frac{\mathcal{W}}{\varphi} \nabla_{\mu} \varphi \nabla_{\nu} \varphi = -8 \pi G T_{\mu \nu}\,,
\label{eq:Einstein}
\end{align}
where $\Box \equiv g^{\mu\nu}\nabla_{\mu}\nabla_{\nu}$ is the d'Alembertian operator \citep[e.g.][]{Quiros+16}.
Varying Eq.~(\ref{eq:TeSaction}) with respect to the scalar field yields the equation for $\varphi$
\begin{equation}
\varphi R+\left(\frac{\mathcal{W}}{\varphi}-\frac{d\mathcal{W}}{d\varphi}\right)\nabla^{\alpha}\varphi \nabla_{\alpha}\varphi-2\mathcal{W}\Box \varphi+2\varphi\frac{d\mathcal{V}}{d\varphi}=0\,.
\label{eq:varphimotion}
\end{equation}

We define CRG by setting
\begin{equation}
  \mathcal{W} (\varphi) = -1\,.
\label{eq:W}
\end{equation}
We note that the original Brans-Dicke theory --- with a constant $\mathcal{W}$ of order unity and a zero potential --- is ruled out by post-Newtonian expansions and solar system experimental tests, which give the constraint $\mathcal{W} \gtrsim 40\,000$ \citep{Faraoni+04, Clifton+12},  and by recent results from CMB observations \citep[e.g.][]{Li2013, Avilez2014}. However, these constraints seem to be weaker on large cosmological scales and can be avoided by adding a self-interaction potential \citep{Hrycyna2014, Quiros2019}, which we define as
\begin{equation} 
  \mathcal{V} (\varphi) = -\Xi \varphi\,,
\label{eq:potential}
\end{equation}
with $\Xi$ being a constant parameter.

With the definitions of Eqs.~(\ref{eq:W}) and (\ref{eq:potential}), the modified field equations and the equation for $\varphi$, in the JF, become
\begin{align}
&\varphi \left(R_{\mu \nu} - \frac{1}{2} g_{\mu \nu} R \right)  - \left(-\frac{1}{2 \varphi} \nabla^{\alpha} \varphi \nabla_{\alpha} \varphi + \Box \varphi - \Xi\varphi\right)g_{\mu \nu} + \nonumber\\
&\quad + \nabla_{\mu} \nabla_{\nu} \varphi - \frac{1}{\varphi} \nabla_{\mu} \varphi \nabla_{\nu} \varphi = -8 \pi G T_{\mu \nu}\,,
\label{eq:Einstein2}\\
	&\varphi R - \frac{1}{\varphi} \nabla^{\alpha} \varphi \nabla_{\alpha} \varphi + 2\Box \varphi - 2\Xi\varphi = 0\,.
\label{eq:varphimotion2}
\end{align}
By using Eq.~(\ref{eq:varphimotion2}) to simplify Eq.~(\ref{eq:Einstein2}), and by using  Eq.~(\ref{eq:Einstein2}), contracted with $g^{\mu \nu}$, to simplify the resulting Eq.~(\ref{eq:varphimotion2}), we obtain the following CRG equations:
\begin{align}
	\varphi R_{\mu \nu} + \nabla_{\mu} \nabla_{\nu} \varphi -\frac{1}{\varphi} \nabla_{\mu} \varphi \nabla_{\nu} \varphi &= -8\pi G T_{\mu \nu}\,,
\label{eq:Einstein3}\\
	\Box \varphi - 2\Xi\varphi &= 8 \pi G T\,. 
\label{eq:varphimotion3}
\end{align}

We derive the stress-energy tensor for the scalar field by recasting Eq.~(\ref{eq:Einstein2}) as 
\begin{equation}\label{modfieldeqscalarsetensor}
\varphi \left(R_{\mu\nu}-\frac{1}{2}g_{\mu\nu}R \right) = -8\pi G\left(T^{\rm M}_{\mu\nu}+T^{\varphi}_{\mu\nu} \right),
\end{equation}
where $T^{\rm M}_{\mu\nu}$ refers to the matter/energy stress-energy tensor, and $T^{\varphi}_{\mu\nu}$ refers to the scalar-field stress-energy tensor, whose explicit form is\footnote{There is actually an ambiguity in the definition of $T^\varphi_{\mu\nu}$, because Eq.~(\ref{eq:Einstein2}) can also be written as $R_{\mu\nu}-\frac{1}{2}g_{\mu\nu}R = -8\pi G \left(T^{\rm M}_{\mu\nu}/\varphi +T^{\varphi,\mathrm{II}}_{\mu\nu}  \right)$, where $T^{\varphi}_{\mu\nu}$ and $T^{\varphi,\mathrm{II}}_{\mu\nu}$ simply differ by a factor $\varphi$. This ambiguity was studied in \citet{Bellucci:2001cc}, where the definition in Eq.~(\ref{modfieldeqscalarsetensor}) is referred to as the effective coupling approach, whereas the latter definition leading to $T^{\varphi,\mathrm{II}}_{\mu\nu}$ is called the mixed approach. Here, we adopt the former definition, which is analogous to the identification of the scalar field $\varphi$ in the Brans-Dicke theory with the inverse of an effective gravitational constant $G_{\rm eff}\sim G/\varphi$.}
\begin{equation}\label{SETensorscalarfield}
-8\pi G T^{\varphi}_{\mu\nu} = g_{\mu\nu}\left(-\frac{1}{2\varphi} \nabla^\alpha \varphi \nabla_\alpha \varphi + \Box \varphi -\Xi \varphi \right)
-\nabla_\mu \nabla_\nu \varphi +\frac{1}{\varphi}\nabla_\mu \varphi \nabla_\nu \varphi\,.
\end{equation}
We use the stress-energy tensor of $\varphi$ to derive an effective dark energy in Sect. \ref{sec:EoSDE}.

\subsection{Relation to Horndeski theories}
\label{sec:Horndeski}

Scalar-tensor theories belong to the wider family of Horndeski models \citep[for reviews on the topic, see, e.g.,][]{Joyce+15, Kobayashi+19}, and, consequently, CRG can also be identified as a special case of Horndeski theories. The detection of GW170817 and GRB170817A \citep{Kase:2018aps, Wang:2017rpx, Sakstein:2017xjx} has greatly constrained the Horndeski theory space. The currently allowed  Lagrangian is \citep[][and references therein]{Noller+19, Creminelli+17, Sakstein+17} 
\begin{equation}
\mathcal{L}_\mathrm{H} = \mathcal{G}_4 \left(\varphi \right) R + \mathcal{G}_2 \left(\varphi, X \right) - \mathcal{G}_3 \left(\varphi, X \right)\Box \varphi\,,
\label{eq:Horndeski}
\end{equation}
where  $\mathcal{G}_i$'s are functions of the scalar field $\varphi$ and $X \equiv \nabla^{\alpha} \varphi \nabla_{\alpha} \varphi$ is the kinetic term. By comparing Eq.~(\ref{eq:Horndeski}) with the Lagrangian of Eq.~(\ref{eq:TeSaction}), and by ignoring the factor $1/(16 \pi G)$, we infer that $\mathcal{G}_4 \leftrightarrow \varphi$, $\mathcal{G}_2 \leftrightarrow -\frac{1}{\varphi} \nabla^{\alpha} \varphi \nabla_{\alpha} \varphi-2\Xi \varphi$, and $\mathcal{G}_3 \leftrightarrow 0$.\footnote{The $f(R)$ models of gravity are also obtained from the Lagrangian of Eq.~(\ref{eq:Horndeski}) by setting $\mathcal{G}_3 = 0$. This result is expected because of the equivalence between the Brans-Dicke theory and $f(R)$ gravity \citep{f(R)Faraoni}.} Therefore, CRG is expected to inherit the important properties of the viable Horndeski theories, including the absence of ghost-like degrees of freedom, namely the Ostrogradski instabilities \citep{Gleyzes+14}.

\subsection{Screening mechanism of CRG}
\label{sec:Screening}

General relativity accurately describes the gravitational interactions on the scale of stars and smaller scales, including the strong gravitational field regime. Therefore, any attempt to modify the theory of gravity by adding new degrees of freedom must provide a screening mechanism to avoid detectable discrepancies in the local tests of gravity \citep{Joyce+15}.

The screening mechanism depends on the local mass density and/or the local gravitational potential. It is thus convenient to study the screening mechanism in the Einstein frame (EF), where the scalar field is minimally coupled to both gravity and the matter fields. The advantage of the EF is that the field equations have a manifestly GR-like form\footnote{Despite their GR-like form, the field equations in the EF are not those of standard gravity. For example, in the vacuum solution of the EF, the scalar field acts as an additional gravitational source. Moreover, the extra coupling between the scalar and matter fields introduces both deviations from the geodesic motions of free-falling particles and a stress-energy tensor which is not covariantly conserved \citep{Faraoni+04,Clifton+12}.} and computations can be performed more easily. However, the presence of the extra coupling between the scalar and matter fields alters rods and clocks, and makes the identification of physical observables harder.

The transition from the JF to the EF can be performed with the conformal transformation of the metric \citep{Clifton+12}, namely a length-scale transformation \footnote{Since a direct measurement of an absolute scale is not possible, experiments are unable to distinguish between the EF and the JF frames. These frames represent two different realizations of the same theory, and physical observables must be equivalent in the two frames \citep[see, e.g.,][]{Sotiriou:2007zu,Postma:2014vaa}.}
\begin{equation}
\tilde{g}_{\mu\nu} = e^{-2\Gamma} g_{\mu\nu}\quad\mathrm{and}\quad\sqrt{\tilde{g}} = e^{-4\Gamma}\sqrt{g}\,,
\label{conftransf}
\end{equation}
where the tilde indicates quantities computed in the EF and  $\Gamma =-\frac{1}{2}\ln \varphi$.
From the JF action of Eq.~(\ref{eq:TeSaction}),  the corresponding action in the EF is 
\begin{align}
\tilde{\mathcal{S}} &= \frac{1}{4\pi G} \int d^4 x \ \sqrt{\tilde{g}} \left[\frac{\tilde{R}}{4}-\frac{1}{2}\tilde{g}^{\mu\nu} \tilde{\nabla}_{\mu} \Psi \tilde{\nabla}_{\nu} \Psi  - \mathcal{U}(\Psi) \right]+ \nonumber\\
&\quad+ \mathcal{S}_\mathrm{m} \left[A^2(\Psi) \tilde{g}_{\mu\nu}; \psi_\mathrm{m}\right]\, ,
\label{ActionEF}
\end{align}
where $\Psi$ is the scalar field related to the scalar field in the JF by $\Psi = \mp \ln \varphi$, $\tilde{R}$ is the Ricci scalar, $A(\Psi) \equiv \varphi^{-1/2}$ is the coupling between the scalar field and matter, and $\mathcal{U}(\Psi)$ is the potential \citep{Clifton+12}
\begin{equation}
\mathcal{U}(\Psi) = 2\Xi e^{\mp \Psi}\,.
\label{eq:UPsiEF}
\end{equation}
 The potential $\mathcal{U}(\Psi) = 2\Xi e^{- \Psi}$ is the runaway potential of the chameleon cosmology \citep{Clifton+12, Joyce+15, Khoury2004}. This potential can be studied by considering the non-relativistic equation of motion for the scalar field \citep[see, e.g.,][and references therein]{Clifton+12}
 \begin{equation}
     \nabla^2 \Psi = \frac{dV_\text{eff}}{d\Psi}\, , \qquad V_\text{eff} \equiv \frac{\Xi}{2\pi G} e^{-\Psi} + \rho e^{\Psi}\, ,
 \end{equation}
 where $V_\text{eff}$ is the effective potential. In particular, $d^2V_\text{eff}/d\Psi^2$ can be interpreted as an effective mass of the scalar field, $m_\Psi$, which reads as
 \begin{equation}
     \frac{d^2V_\text{eff}}{d\Psi^2}= m_\Psi = \frac{\Xi}{2\pi G} e^{-\Psi} + \rho e^{\Psi}\, .
     \label{mPsi}
 \end{equation}
In regimes of large densities $\rho \gg e^{-2\Psi} \Xi/2\pi G$, the scalar field becomes increasingly more massive and hence the fifth force mediated by it has an increasingly shorter range; in other words, in this regime the presence of the scalar field is effectively screened. 
Conversely, in regimes of small densities $\rho \ll e^{-2\Psi}\Xi/2\pi G$, $m_\Psi$ gets smaller, the force mediated by $\Psi$ has an increasingly longer range, and the field is free to propagate.\footnote{ A theorem guarantees that in TeS theories endowed with a potential $\mathcal{U}$ that satisfies the condition $d^2\mathcal{U}(\Psi)/d\Psi^2 >0$, like CRG (Eq.~\ref{eq:UPsiEF}),  black holes in vacuum are equivalent to GR black holes 
\citep{Bekenstein:1995un,Sotiriou:2011dz,Cruz:2017ecg}. 
In CRG, black holes embedded in environment with density sufficiently small to make the screening mechanism ineffective, might in principle develop scalar hair. 
However, in extended theories of gravity, black holes or compact objects with scalar hair  remain viable and their existence can be tested with gravitational wave observations \citep{Sotiriou:2015pka,Berti:2015itd,Cardoso:2016oxy,Brito:2017zvb,Barack:2018yly,Maggiore:2019uih}.}
The  density scale separating the two regimes thus depends on  both  $\Psi$ and  $\Xi$.  This result is consistent with the assumption, in the phenomenological RG, that the gravitational permittivity depends on the local density, and  suggests a relation between $\epsilon$ and $\Psi$, or equivalently $\varphi$. In addition, this result indicates that the density scale depends on $\Xi$, which is a constant universal value independent of the local environment. In Sect. \ref{sec:EoSDE} and Appendix \ref{app:SNIa}, we find that $\Xi$ plays the role of the cosmological constant $\Lambda$ in the standard cosmological model. Therefore, we expect that the local value of the scalar field $\Psi$ (or $\varphi$), rather than $\Xi$, plays the major role in setting the density threshold for the screening mechanism.


\subsection{The weak-field limit of CRG}
\label{sec:weakfieldlimit}

As in standard GR, we take the static weak-field limit of the metric $g_{\mu \nu}$,\footnote{We adopt the signature $(-,+,+,+)$ for the Minkowski metric.} which can be expanded around the Minkowski metric $\eta_{\mu \nu}$ as
\begin{equation}
g_{\mu \nu} \simeq \eta_{\mu \nu} + h_{\mu \nu}\,,
\end{equation}
where $|h| \ll 1$ and $|\partial_\mu h| \ll 1$. We write the metric as
\begin{align}
g_{00} &\simeq \eta_{00} - 2\Phi\,,
\label{eq:weakfield-g00}\\
g_{0i} &\simeq 0\,, \\
g_{ij} &\simeq \eta_{ij} - 2U\,,
\label{eq:weakfield-gij}
\end{align}
where $\Phi$ and $U$ are two potentials, and we ignore terms of order $\mathcal{O}(\Phi^2) \sim \mathcal{O}(U^2)$. These expressions enable the calculation of the left-hand side of Eqs.~(\ref{eq:Einstein3}) and (\ref{eq:varphimotion3}) up to terms of order $\mathcal{O}(\Phi) \sim \mathcal{O}(U)$ (see Appendix~\ref{app:weak_field} for the detailed calculations). 

For the right-hand side of Eqs.~(\ref{eq:Einstein3}) and (\ref{eq:varphimotion3}), we consider a static non-relativistic fluid: the only non-zero component of the energy-momentum tensor is $T_{00} \simeq \rho$, and thus its trace is $T \simeq -\rho$. With these assumptions, the 00-component of the field equations reduces to
\begin{equation}
\vec{\nabla}\cdot\left(\varphi \vec{\nabla}\Phi\right) \simeq 8\pi G\rho\,.
\label{eq:weak_field_poisson}
\end{equation}
We thus recover the RG equation if we identify the scalar field with twice the permittivity: $\varphi = 2\epsilon$. In the Newtonian regime, we have a constant scalar field, namely $\vec{\nabla}\varphi = 0$, and thus we recover the standard Poisson equation for $\varphi = 2$.

\subsubsection{The gravitational field of a spherical source immersed in a homogeneous background}
\label{Pointvacuumnewtonianlimit}

In Appendix~\ref{WF-SphSource}, we compute the gravitational field generated by a spherical source immersed in a homogeneous background with density $\rho_{\rm bg}$. The source is described by a density profile $\rho_\mathrm{s}(r)$ decreasing with $r$. We estimate the field in the limit  $\rho_\mathrm{s}(r)\gg \rho_\mathrm{bg}$, close to the source, and $\rho_\mathrm{s}(r)\ll \rho_\mathrm{bg}$, at large distance from the source.

At small distances, we find
\begin{equation}
\frac{d\Phi}{dr}
  = \frac{2}{\varphi}\frac{Gm(<r)}{r^2}\,,
\label{eq:PhiCloseSource}
\end{equation}
with the scalar field
\begin{equation}
\varphi(r) = 2\left[1 - \int_0^r\frac{Gm(<r^\prime)}{r'^2}\,dr^\prime\right]\,.
\label{eq:varphiCloseSource}
\end{equation}

At large distances from the source, the scalar field $\varphi$ and the gravitational field $d\Phi/dr$ satisfy the implicit relation 
\begin{equation}
\frac{d\ln\varphi}{dr}
  =
  \frac{d\Phi}{dr}\left\{
  -1
  - \left[
    1 + \left(\frac{d\Phi}{dr}\right)^{-2}
      \left(2\Xi - \frac{8\pi G\rho}{\varphi}\right)\right]^{1/2}\right\}\, ,
\label{eq:PhiDistantSource}
\end{equation}
with $\rho(r)=\rho_\mathrm{s}(r)+\rho_\mathrm{bg}$. This relation sets the acceleration scale
\begin{equation}
 a_\Xi = \left(2\Xi-\frac{8\pi G\rho}{\varphi}\right)^{1/2} \, .
\label{eq:aXi}
\end{equation}
In regions where $d\Phi/dr\gg a_\Xi$, the gravitational field $d\Phi/dr\simeq -(1/2)d\ln\varphi/dr$ has a similar dependence of the field at small distances (Eqs.~\ref{eq:PhiCloseSource} and \ref{eq:varphiCloseSource}).
In regions where $d\Phi/dr\ll a_\Xi$, the RG acceleration deviates from the Newtonian acceleration.

This result resembles the starting hypothesis of MOND, that introduces the acceleration scale $a_0$ to separate the Newtonian from the modified gravity regimes. Moreover, in Sect.~\ref{Sec:RRGCosmology}, we find $\Xi \sim \Lambda$. We thus find that $a_\Xi \sim 10^{-10}$~m~s$^{-2}$, in the limit $2\Xi\gg 8\pi G\rho/\varphi$, that occurs at large distances from the source. The existence of this acceleration scale appears in a number of observations on the scale of galaxies \citep[see, e.g.,][]{deMartino:2020gfi, McGaugh:2020ppt, Chae:2020vlk, McGaugh:2016leg, McGaugh:2018, Merritt:2020pwe}. Indeed, the dynamics of disk and elliptical galaxies in the low-acceleration regime is described in the RG framework without requiring the existence of dark matter \citep{Cesare+20,Cesare+21}. Nevertheless, some of the phenomenology predicted by MOND in the low-acceleration regime, set by an acceleration scale $a_0$ independent of the source, like the radial acceleration relation \citep{McGaugh:2016leg}, appears to be inconsistent with the rotation curves of dwarf disk spirals and low-surface brightness galaxies \citep{DiPaolo:2018mae, Santos-Santos:2019vrw}. These tensions might suggest that the acceleration scale could indeed depend on the source, as it happens for $a_\Xi$ in CRG.  

The connection between $a_\Xi$ and $\Xi$, in the limit $2\Xi\gg 8\pi G\rho/\varphi$, is similar to the connection between $a_0$ and $\Lambda$ in MOND: $a_0 \sim \Lambda^{1/2}$. A number of different  sensible arguments have been suggested for the interpretation of the latter relation  \citep{Milgrom:1989alt, Milgrom:1998sy, Famaey+12, Milgrom:2020cch}. In the CRG context, $a_\Xi\sim (2\Xi)^{1/2}$ emerges naturally. 

\subsubsection{The scalar field $\varphi$ for a spherical source immersed in a constant background}

According to the results of Appendix~\ref{WF-SphSource}, the scalar field $\varphi$ is positive and broadly in the range $[0,\,2]$ consistently with the RG {\it ansatz} $\epsilon \in [0, 1]$ for the permittivity. In the limit $\rho_\mathrm{bg}\ll \rho_\mathrm{s}$, we have 
\begin{equation}
\varphi_\mathrm{near}
  \simeq 2\left[1 - \int_0^r\frac{Gm(<r^\prime)}{r'^2}\,dr^\prime \right]\, , 
\end{equation}
whereas in the limit $\rho_\mathrm{bg}\gg \rho_\mathrm{s}$,
\begin{equation}
\varphi_\mathrm{far} \simeq \frac{4\pi G}{\Xi}\left(\rho_\mathrm{bg}
  + \rho_\mathrm{s}\right)\,.
\end{equation}
These limits are broadly in agreement with the smooth step function considered for $\epsilon = \epsilon(\rho)$ in previous RG studies \citep{Cesare+21,Cesare+20,MatsakosDiaferio16}. In those studies, the local mass density was found to be a good proxy of the transition between the Newtonian and the RG regimes.

\section{The homogeneous and isotropic universe in CRG}
\label{Sec:RRGCosmology}

We derive the basic equations of a homogeneous and isotropic universe in Sect. \ref{sec:basiceq} and we  solve these equations for a spatially flat universe in Sect. \ref{sec:flat}. We derive the equation of state of the effective dark energy in Sect. \ref{sec:EoSDE} and discuss the evolution of the scalar field in Sect. \ref{sec:evolutionphi}.

\subsection{Basic equations}
\label{sec:basiceq}

The covariant formulation of RG enables the description of a homogeneous and isotropic universe that can be described by the Friedmann-Lema\^{\i}tre-Robertson-Walker (FLRW) metric
\begin{equation}
ds^2 = -dt^2 + a(t)^2\left(\frac{dr^2}{1-kr^2} + r^2 d\theta^2 + r^2 \sin^2 \theta \ d\phi^2 \right),
\label{eq:FLRWmetric}
\end{equation}
where $a(t)$ is the scale factor, $k$ is the spatial curvature, and $(t, r, \theta, \phi)$ are the co-moving coordinates. By treating the content of the Universe as a perfect fluid, Eq.~(\ref{eq:Einstein3}) readily yields the modified Friedmann equations (see Appendix~\ref{app:friedmann})
\begin{equation} 
	 \frac{\ddot{a}}{a} - \frac{1}{3}\left(\frac{\dot{\varphi}^2}{\varphi^2}-\frac{\ddot{\varphi}}{\varphi} \right) = -\frac{8\pi G}{3\varphi}\rho
	\label{eq:friedmann1}
\end{equation}
and
\begin{equation} 
	 \frac{\ddot{a}}{a} + 2\frac{\dot{a}^2}{a^2} + \frac{2k}{a^2} + \frac{\dot{a}\dot{\varphi}}{a\varphi} = \frac{8\pi G}{\varphi} p\,.
	\label{eq:friedmann2}
\end{equation}
The equation for the scalar field, Eq.~(\ref{eq:varphimotion3}), reduces to (see Appendix~\ref{app:friedmann})
\begin{equation}
	\frac{1}{3}\frac{\ddot{\varphi}}{\varphi} + \frac{\dot{a}\dot{\varphi}}{a\varphi} + \frac{2\Xi}{3} = \frac{8\pi G}{3\varphi} \left(\rho - 3p \right)\,.
\label{scalarfieldcosmo}
\end{equation}

In the JF, the stress-energy tensor is covariantly conserved \citep{Faraoni+04}. We assume the standard equation of state 
\begin{equation}
p = w \rho \,,
\label{EOS}
\end{equation}
with $w = 0$ and $w = 1/3$ describing the dust and radiation components, respectively.
The dependence of the matter or radiation density on time $t$, or equivalently on the scale factor $a(t)$, is thus 
\begin{equation}
\rho(t) =  \rho_0  a^{-3\left(w+1 \right)} \, ,
	\label{eq:density}
\end{equation}
with $\rho_0$ the mean density of the component at the present time, $t_0$, when $a(t_0) = a_0 = 1$.

\subsection{A spatially flat universe: Analytic solution}
\label{sec:flat}

Here, we solve the field equations, Eqs.~(\ref{eq:friedmann1})-(\ref{scalarfieldcosmo}), in the special case of a spatially flat universe with $k=0$. We assume that the universe only contains baryonic matter with density $\rho_{\rm b}$ and negligible pressure, $p=0$, namely $w=0$ in the equation of state, Eq.~(\ref{EOS}). 

After substituting Eq.~(\ref{eq:friedmann1}) into Eq.~(\ref{eq:friedmann2}) and dividing by $H^2 \equiv (\dot{a}/a)^2$, Eqs.~(\ref{eq:friedmann1})-(\ref{scalarfieldcosmo}) become
\begin{align}
\frac{1}{H^2}\frac{\ddot{a}}{a}-\frac{1}{3H^2}\left(\frac{\dot{\varphi}^2}{\varphi^2}-\frac{\ddot{\varphi}}{\varphi} \right) &= -\frac{8\pi G}{3H^2\varphi}\rho_{\rm b}\,, \label{ModFried00} \\
2 + \frac{1}{H}\frac{\dot{\varphi}}{\varphi } + \frac{1}{3H^2}\left(\frac{\dot{\varphi}^2}{\varphi^2}-\frac{\ddot{\varphi}}{\varphi} \right) &= \frac{8\pi G}{3H^2\varphi}\rho_{\rm b}\,, \label{ModFriedrr}\\
\frac{1}{3H^2}\frac{\ddot{\varphi}}{\varphi} + \frac{1}{H} \frac{\dot{\varphi}}{\varphi } +\frac{2}{3H^2}\Xi &= \frac{8\pi G}{3H^2\varphi} \rho_{\rm b}\,. \label{ScalarCosmo}
\end{align}
The solution of the above system of coupled equations together with the equation of the mass-energy conservation, Eq.~(\ref{eq:density}), determines the time evolution of both the scalar field and the scale factor.

We simplify the field equations by introducing the modified cosmological parameter
\begin{equation}
\Omega \equiv \frac{2\Omega_\mathrm{b}}{\varphi} \equiv \frac{16\pi G}{3 H^2\varphi} \rho_{\rm b}\, , 
\label{cosmparamOm}
\end{equation}
where  $\Omega_\mathrm{b}=\rho_{\rm b}/\rho_{\rm cr}\equiv \rho_{\rm b} (8\pi G/3 H^2)$ is the density parameter associated with the homogeneous baryonic density $\rho_{\rm b}$. The density parameter $\Omega$ is analogous to the total matter density parameter of the standard model, that includes both baryonic and dark matter; in CRG, the gravitational role of the baryonic matter density is amplified by the factor $2/\varphi$, unlike the standard model, where the baryonic and the dark matter components simply add up.

Similarly, we define
\begin{equation}
 \Omega_{\Xi} = \frac{\Xi}{3H^2}\, .
\label{cosmparamOl}
\end{equation}
The parameter $\Omega_\Xi$ is analogous to the standard cosmological parameter $\Omega_\Lambda=\Lambda/3H^2$, with $\Lambda$ the cosmological constant in the GR field equations; therefore, in CRG, $\Xi$ exactly plays the role of $\Lambda$ in the standard model.   At $t=t_0$, the values of the two parameters $\Omega$ and $\Omega_\Xi$ are $\Omega_0 = 16\pi G\rho_{\rm b0}/3H_0^2\varphi_0$ and $\Omega_{\Xi0} = \Xi/3H_0^2$. 

We further introduce two deceleration parameters, $q$ and $q_\varphi$, related to the scale factor and the scalar field, respectively:
\begin{equation}
q \equiv -\frac{a \ddot{a}}{\dot{a}^2} = -\frac{\ddot{a}}{aH^2}\,, \quad q_{\varphi} \equiv -\frac{\ddot{\varphi}}{\varphi H^2}\,.
\label{decparameters}
\end{equation}
With the above definitions, the field equations, Eqs.~(\ref{ModFried00})-(\ref{ScalarCosmo}), become
\begin{align}
-q-\frac{1}{3H^2}\left(\frac{\dot{\varphi}}{\varphi} \right)^2- \frac{1}{3}q_{\varphi} &= -\frac{1}{2}\Omega\,,
\label{system3a}\\
2+ \frac{1}{H}\frac{\dot{\varphi}}{\varphi } + \frac{1}{3H^2}\left(\frac{\dot{\varphi}}{\varphi}\right)^2+\frac{1}{3} q_{\varphi} &= \frac{1}{2}\Omega\,,
\label{system3b}\\
-\frac{1}{3}q_{\varphi}+ \frac{1}{H} \frac{\dot{\varphi}}{\varphi } + 2\Omega_{\Xi} &= \frac{1}{2}\Omega\,.
\label{system3c}
\end{align}
It is convenient to define 
the new quantity
\begin{equation}
\zeta \equiv \frac{\dot{\varphi}}{H \varphi}\,.
\label{psi}
\end{equation}
The field equations, Eqs.~(\ref{system3a})-(\ref{system3c}), now become 
\begin{align}
q+\frac{1}{3}\zeta^2 +  \frac{1}{3}q_{\varphi} &= \frac{1}{2}\Omega\,,  \label{system4a}\\
2+ \zeta + \frac{1}{3}\zeta^2+\frac{1}{3} q_{\varphi} &= \frac{1}{2}\Omega\,, \label{system4b}\\
-\frac{1}{3}q_{\varphi}+ \zeta &= \frac{1}{2}\Omega - 2\Omega_{\Xi}\, \label{system4c}.
\end{align}
Combining the first two equations yields
\begin{equation}
q-\zeta = 2\,,
\end{equation}
namely, with the definitions of $q$ and $\zeta$,
\begin{equation}
\frac{d}{dt}\ln\varphi = - \frac{d}{dt}\left(\ln\dot{a} + 2\ln{a} \right)\,.
\label{diff1}
\end{equation}
This equation can be integrated to obtain $\varphi = \mathcal{C}/(a^2\dot{a})$, with $\mathcal{C}$ an integration constant. Its value can be found from the boundary conditions $a(t_0)=1$, $H(t_0) = H_0$, and $\varphi(t_0) = \varphi_0$. We obtain  $\mathcal{C} = H_0\varphi_0$ and thus
\begin{equation}
\varphi = \frac{H_0\varphi_0}{Ha^3}\,.
\label{avarphi}
\end{equation}
Moreover, since $3Ha^3 = da^3/dt$, we can write the scalar field as
\begin{equation}
\varphi = 3H_0\varphi_0 \left(\frac{da^3}{dt}\right)^{-1}\,.
\label{avarphi2}
\end{equation}
This expression shows that the scalar field  $\varphi$ is inversely proportional to the rate of the variation of  the volume of the universe.

As shown in Appendix~\ref{app:dHdt}, adding up Eqs.~(\ref{system4a}) and (\ref{system4c}) yields two solutions for $dH/dt$:
\begin{equation}
\frac{dH}{dt} = \pm\sqrt{3}\left(H^4 + \Omega_0H_0H^3 - 2\Omega_{\Xi0}H_0^2H^2\right)^{1/2}\,,
\label{eq:dHdt}
\end{equation}
each solution being a first order differential equation for $H$. Hereafter, we refer to the solution corresponding to the upper ($+$) sign as CRG+, and to the lower ($-$) sign as CRG--, respectively. The integration of Eq.~(\ref{eq:dHdt}) yields
\begin{equation}
H(t)
=\frac{4\Omega_{\Xi0}H_0}
  {\Omega_0
    \mp\left(\Omega_0^2 + 8\Omega_{\Xi0}\right)^{1/2}\sin\left[\sqrt{6\Omega_{\Xi0}}(H_0t + C_1)\right]}\,,
\label{eq:H}
\end{equation}
with the upper and lower signs corresponding to 
CRG+ and CRG--, respectively. 
The integration constant, $C_1$, can be found from the boundary condition at $t=t_0$:
\begin{equation}
C_1 = - H_0t_0
\pm\frac{1}{\sqrt{6\Omega_{\Xi0}}}\arcsin\left[\frac{\Omega_0 - 4\Omega_{\Xi0}}{
  \left(\Omega_0^2 + 8\Omega_{\Xi0}\right)^{1/2}}\right]\,.
\label{eq:C1}
\end{equation}

Furthermore, the scale factor can be derived by integrating Eq.~(\ref{eq:H})
\begin{align}
&a(t)
= \nonumber\\
&C_2\left(\left|
  \frac{\Omega_0
    \tan \tau
  \mp \left(\Omega_0^2 + 8\Omega_{\Xi0}\right)^{1/2}
  - (8\Omega_{\Xi0})^{1/2}}{\Omega_0
    \tan \tau
  \mp \left(\Omega_0^2 + 8\Omega_{\Xi0}\right)^{1/2}
  + (8\Omega_{\Xi0})^{1/2}
}\right|\right)^{1/\sqrt{3}}\,,
\label{eq:scalefactorsol}
\end{align}
where
\begin{equation}
\tau = \left(\frac{3}{2}\Omega_{\Xi0}\right)^{1/2}(H_0t + C_1)\,,
\end{equation}
and $C_2$ is an integration constant, whose value can be fixed by the condition $a_0 = 1$:
\begin{equation}
C_2
= \left(\left|
  \frac{\Omega_0
    \tan\tau_0
  \mp \left(\Omega_0^2 + 8\Omega_{\Xi0}\right)^{1/2}
  + (8\Omega_{\Xi0})^{1/2}}{\Omega_0
    \tan\tau_0
  \mp \left(\Omega_0^2 + 8\Omega_{\Xi0}\right)^{1/2}
  - (8\Omega_{\Xi0})^{1/2}
}\right|\right)^{1/\sqrt{3}}\,,
\end{equation}
with
\begin{equation}
\tau_0 = \pm\frac{1}{2}\arcsin\left[\frac{\Omega_0 - 4\Omega_{\Xi0}}{
  \left(\Omega_0^2 + 8\Omega_{\Xi0}\right)^{1/2}}\right]\,.
\label{eq:tau0}
\end{equation}

Inverting Eq.~(\ref{eq:tau0}) yields
\begin{equation}
\sin\left(\pm 2\tau_0 \right) = \frac{\Omega_0-4\Omega_{\Xi0}}{\left(\Omega_0^2+8\Omega_{\Xi0} \right)^{1/2}},
\end{equation}
which implies that the cosmological parameters $\Omega_0$ and $\Omega_{\Xi0}$ satisfy the relation
\begin{equation}
-1 \le \frac{\Omega_0-4\Omega_{\Xi0}}{\left(\Omega_0^2+8\Omega_{\Xi0} \right)^{1/2}} \le 1.
\label{eq:cosmparambound}
\end{equation}

The parameter $\Omega_0$ must be positive, because it is proportional to the mass-energy density $\rho_{\rm b0}$, according to Eq.~(\ref{cosmparamOm}); therefore, the only physically viable solutions of these inequalities are
\begin{equation}
\Omega_0 >0; \qquad 0 < \Omega_{\Xi0} \le \frac{\Omega_0+1}{2}\, .
\label{eq:cosmparambound2}
\end{equation}
The parameter $\Xi$ must thus be positive.

The first test of the viability of CRG is its ability to properly describe the Hubble diagram of the observed SNeIa at high redshift. Using the scale factor in Eq.~(\ref{eq:scalefactorsol}), the luminosity distance can be computed exactly, for both signs appearing in Eq.~(\ref{eq:scalefactorsol}) (see Appendix~\ref{sec:LumDistAnalytic}). As in the standard cosmological model, the values of the cosmological parameters can be set by modelling the SNIa Hubble diagram  (see Appendix~\ref{app:SNIa}). 

\subsection{The equation of state of the effective dark energy}
\label{sec:EoSDE}

Equation~(\ref{SETensorscalarfield}) is the general expression of the stress-energy tensor of the scalar field. By using the FLRW metric of Eq.~(\ref{eq:FLRWmetric}), Eq. (\ref{SETensorscalarfield}) becomes
\begin{equation}\label{cosmologicalSETscalarfield}
\begin{split}
T^{\varphi}_{\mu\nu}= &-\frac{g_{\mu\nu}}{8\pi G}\left(\frac{\dot \varphi^2}{2\varphi}-\ddot \varphi-3H\dot \varphi-\Xi \varphi\right)+\\
&+\frac{1}{8\pi G}\nabla_\mu \nabla_\nu \varphi -\frac{1}{8\pi G \varphi}\nabla_\mu \varphi \nabla_\nu \varphi\,.
\end{split}
\end{equation}
If we consider $T^{\varphi}_{\mu\nu}$ to be in the form of a perfect fluid, $T^{\varphi}_{\mu\nu} = \left(\rho_\varphi + p_\varphi \right)u_\mu u_\nu + p_\varphi g_{\mu\nu}$, the $00$- and $ij$-components of Eq.~(\ref{cosmologicalSETscalarfield}) lead to the effective density and pressure associated with the scalar field:
\begin{align}
\rho_\varphi &= -\frac{1}{8\pi G}\left(\frac{\dot\varphi^2}{2\varphi}+3H\dot\varphi+\Xi \varphi \right)\,, \label{rhovarphi}\\
p_\varphi &= -\frac{1}{8\pi G}\left(\frac{\dot\varphi^2}{2\varphi}-\ddot\varphi-2H\dot\varphi-\Xi \varphi \right)\,.\label{pvarphi}
\end{align}
In order to derive the density and pressure associated with the effective dark energy in CRG, we can combine Eq.~(\ref{eq:friedmann1}) with Eq.~(\ref{scalarfieldcosmo}), and Eq.~(\ref{eq:friedmann2}) with Eq.~(\ref{scalarfieldcosmo}), to rewrite the modified Friedmann Eqs.~(\ref{eq:friedmann1})-(\ref{eq:friedmann2}) as
\begin{align}
\ddot{a} &= -\frac{4\pi G}{3\varphi}\left(\rho + 3p+ \rho_\varphi + 3p_\varphi \right)a\,,\\
\dot{a}^2 &= \frac{8\pi G}{3\varphi}\left(\rho + \rho_\varphi \right)a^2\,.
\end{align}

By comparing these equations with those of a general TeS theory with a non-minimal coupling between the scalar field and the metric,
\begin{align}
\ddot{a} &= -\frac{4\pi G}{3\varphi_0}\left(\rho + 3p+ \rho_{\rm DE}+ 3p_{\rm DE}\right)a\,, \label{FriedmannDE1}\\
\dot{a}^2 &= \frac{8\pi G}{3\varphi_0}\left(\rho + \rho_{\rm DE}\right)a^2\,, \label{FriedmannDE2}
\end{align}
where $\varphi_0$ is the value of the scalar field computed at the present epoch,\footnote{In standard TeS theories, $\varphi$ is often regarded as the inverse of the gravitational constant, justifying the presence of the term $1/\varphi_0$ in Eqs.~(\ref{FriedmannDE1})-(\ref{FriedmannDE2}). Based on our convention here, the scalar field is adimensional and the effective gravitational constant is $G/\varphi$.}
we derive the general expressions
\begin{align}
\rho_{\rm DE} &= \frac{\varphi_0}{\varphi}\rho_\varphi + \left(\frac{\varphi_0}{\varphi}-1 \right)\rho\,, \label{rhoDEeff}\\
p_{\rm DE} &= \frac{\varphi_0}{\varphi}p_\varphi + \left(\frac{\varphi_0}{\varphi}-1 \right)p\,, \label{pDEeff}
\end{align}
where $\rho = \rho_{\rm m} + \rho_{\rm r}$ and $p = p_{\rm r}=\rho_{\rm r}/3$ are the density and pressure associated with matter, $\rho_{\rm m}$, and radiation, $\rho_{\rm r}$ and $p_{\rm r}$, respectively \citep{Frusciante:2019xia}.

Based on these definitions, we can evaluate the parameter associated with the equation of state of the effective dark energy, $w_{\rm DE}=p_{\rm DE}/\rho_{\rm DE}$. Its analytical expression is reported in Eq.~(\ref{eq:wDE}) of Appendix~\ref{sec:EoSEffDE}. 
At the present epoch, for $H = H_0$ and $a = a_0= 1$, we obtain 
\begin{equation}\label{wDEpresent}
w_\mathrm{DE} =
    -\frac{6 \pm 4\sqrt{3}\left(1 + \Omega_0 - 2\Omega_{\Xi0}\right)^{1/2}}
    {6 - 3\Omega_0}\,,
\end{equation}
where the upper and lower signs refer to the CRG+ and the CRG-- solutions, respectively.

The value which best fits the observational data is $w_{\rm DE} = -1$ \citep{Aghanim:2018eyx}, which is consistent with the cosmological constant of the $\Lambda$CDM model. By assuming $\Omega_0\sim 0.3$, we find $\Omega_{\Xi0} \sim 0.64$, which is only admitted by CRG-- and is near the upper bound of Eq.~(\ref{eq:cosmparambound2}). 

Values of $w_{\rm DE}$ slightly different from $-1$ can nonetheless be accommodated by the data \citep[e.g.][]{Amendola:2015ksp,Copeland:2006wr,Frusciante:2019xia,Wen:2017aaa,Capozziello:2005mj,Gerardi:2019obr}. 
Observational constraints on the equation of state of the effective dark energy generally depend on the model used to describe its effects \citep[but see][for a model-independent reconstruction]{Gerardi:2019obr}. We can gain some insight by adopting 
the parametrization 
$w_{\rm DE}(z) = w_0 + w_a z/(1+z)$ \citep{Chevallier:2000qy, Linder:2002et}. At the present epoch $z=0$, this parametrization only depends on $w_0$, and it is therefore sufficiently general to account for a broad range of dark energy models, either with $w_{\rm DE}<-1$ (phantom models) or $w_{\rm DE}\geq-1$ \citep{Amendola:2015ksp,Frusciante:2019xia,Copeland:2006wr,Sola:2005et,Bean:2001xy}. Baryonic acoustic oscillations (BAO), SNIa and CMB observational data \citep{Hazra:2013dsx,Wen:2017aaa} constrain $w_0$ to be approximately in the range $w_0 \in [-1.18,\, -0.85]$. 
For $\Omega_0 = 0.3$, this range of $w_0$ yields, by using Eq.~(\ref{wDEpresent}), $\Omega_{\Xi0}\in [0.62,\,0.65]$ fot the CGR-- solution; for the  CRG+ solution, the allowed range of $w_0$ is limited to $w_0<-1.1765$ and it thus only admits $\Omega_{\Xi0}\simeq 0.65$. 

 These values of $\Omega_{\Xi0}$ close to the upper limit $(1+\Omega_0)/2$ (Eq. \ref{eq:cosmparambound2}) are consistent with the constraints we obtain from the SNIa data that we investigate in Appendix \ref{app:SNIa}.
Future observations aimed at investigating the large-scale structure of the Universe and the evolution of dark energy \citep{Amendola:2016saw} will further constrain $w_0$, and hence the value of $\Xi$. 

\subsection{Additional remarks on the evolution of the scalar field}
\label{sec:evolutionphi}

By using Eqs.~(\ref{cosmparamOm}), (\ref{cosmparamOl}), and (\ref{eq:density}) for a matter-dominated universe ($p = 0$), we can write the field equation of the scalar field, Eq.~(\ref{avarphi}), as
\begin{equation}
\frac{1}{3}\frac{\ddot{\varphi}}{\varphi} + H\frac{\dot{\varphi}}{\varphi}
  = \left(\frac{H}{H_0}\frac{\Omega_0}{2} - 2\Omega_{\Xi0}\right)H_0^2\,.
\end{equation}
The right-hand side of this equation is negative  at the present time $t_0$, for $\Omega_0 = 0.3$ and $\Omega_{\Xi0} = 0.65$. In other words, the contribution of $\Omega_{\Xi 0}$ dominates over $\Omega_0$. However, the $\Xi$ term plays a decreasingly important role at increasing redshifts with 
the right-hand side of the equation above becoming positive, by adopting $H(t)$ from Eq.~(\ref{eq:H}), at times $t$ earlier than
\begin{equation}
t \simeq t_0 - \frac{\pi}{4H_0}\, .
\end{equation}
In the radiation-dominated era, however, the right-hand side of Eq.~(\ref{scalarfieldcosmo}) is zero, and therefore $\Xi$ may become an important source term in the scalar field equation.

Big Bang nucleosynthesis could potentially constrain the value of $\Xi$ and, more generally, the scalar potential. Scalar-tensor theories, and  specifically Brans-Dicke-like models, provide a time-varying effective gravitational constant. This feature alters the rate of expansion of the universe and therefore its temperature, which both contribute to regulate the primordial nucleosynthesis. In particular, the observed abundances of $^2$H, $^3$He, and $^7$Li set a bound on the present matter density, whereas the abundance of $^4$He constrains the rate of change of $\varphi$ \citep{arai1987primordial}. For TeS theories without a self-interaction potential, there is an attractor mechanism towards GR \citep{Damour:1998ae} as the scalar field, and thus $G$, remains approximately constant during the radiation-dominated epoch. Therefore, in these models, the nucleosynthesis of light elements occurs similarly to the standard cosmological scenario, but with a different expansion rate \citep{Damour:1990eh,Casas:1990fz,Casas:1991ui,serna1992primordial,Clifton:2005xr}. The primordial nucleosynthesis in the presence of a self-interaction potential --- as in CRG --- has been extensively studied in the literature \citep[see, e.g.,][and references therein]{arai1987primordial,Uzan:2002vq,Larena:2005tu,Coc:2006rt,Iocco:2008va,Clifton+12}, also as a possible solution to the problem of the $^7$Li abundance. An extensive analysis of whether CRG is consistent with the observed abundances of light elements is left to future work.

\section{Conclusion}
\label{Sec:Conclusions}

Refracted gravity was originally introduced to describe the dynamics of galaxies and galaxy systems without the aid of dark matter. The contribution of dark matter to the gravitational field is mimicked by the gravitational permittivity $\epsilon(\rho)$, a monotonic function of the mass density $\rho$ \citep{MatsakosDiaferio16}. 
Here, we propose a covariant extension of RG, CRG, and show that it belongs to the family of scalar-tensor theories, thus inheriting all their general properties. A scalar-tensor theory is specified by the self-interacting potential $\mathcal{V}(\varphi)$ of the scalar field $\varphi$ and by the general differentiable function $\mathcal{W}(\varphi)$ appearing in the Lagrangian density. For CRG, we adopt $\mathcal{V}(\varphi)=-\Xi\varphi$, with $\Xi$ a normalization constant, and $\mathcal{W}(\varphi)=-1$. In the weak-field limit, this theory correctly reduces to the original phenomenological RG and identifies the gravitational permittivity with the scalar field, with $\varphi=2\epsilon$. 

When used to describe an expanding and homogeneous  universe, the scalar field $\varphi$ is also responsible for the observed accelerated expansion of the Universe. The cosmological density associated with the cosmological constant in the standard model, $\Omega_{\Lambda 0}=\Lambda/3H_0^2$, is now replaced by the CRG parameter $\Omega_{\Xi 0}=\Xi/3H_0^2$. It follows that the normalization constant $\Xi$ plays the role of the cosmological constant $\Lambda$ of the standard model. Moreover, the Hubble diagram of high-redshift SNeIa and current observational constraints on the parameter $w_\mathrm{DE}$ of the equation of state of the effective dark energy, that we derive from the stress-energy tensor associated with the scalar field, suggest that,  in a universe with a flat geometry,  $\Omega_{\Xi0}\sim (1+\Omega_0)/2$, with $\Omega_0$ the ratio between the baryonic matter and the homogeneous scalar field at the present time. The parameter $\Omega_0$ can be identified with $\Omega_0=0.31$ of the standard model \citep{Aghanim:2018eyx}. We find $\Omega_{\Xi0}=0.650^{+0.005}_{-0.085}$ at the 90\% confidence level for the CRG+ solution. It thus follows that $\Xi$ and $\Lambda$ have comparable values. 

In addition, in the weak-field limit, $\Xi$ sets an acceleration scale, $(2\Xi)^{1/2}\sim 10^{-10}$~m~s$^{-2}$, below which RG deviates from Newtonian gravity and appears to describe the dynamics of disk and elliptical galaxies without the aid of dark matter \citep{Cesare+20, Cesare+21}. This acceleration scale is indeed present in real systems \citep{Chae:2020vlk, McGaugh:2018} and is comparable to the acceleration $a_0$ introduced in MOND \citep{Milgrom1, Milgrom2, Milgrom3}. Therefore, 
being $\Xi\sim \Lambda$, the known relation $a_0\sim \Lambda^{1/2}$ \citep{Milgrom:1998sy, Milgrom:2020cch} naturally appears in CRG.

CRG provides a connection between phenomena generally attributed to dark matter and dark energy separately and it thus belongs to the family of modified gravity models that connect 
the two dark sectors within a unified scenario. 
The same property is currently shared by other models, including some $f(R)$ theories \citep{f(R)Faraoni}, ``quartessence'' theories \citep{Brandenberger+19} or generalized Chaplygin gas models \citep{Bento:2002ps,Zhang:2004gc}. 

Assessing if the evolution of the universe described by CRG is consistent with observations requires extensive additional work. For example, the power spectrum of the temperature anisotropies of the CMB and the power spectrum of matter density perturbations are required to constrain the Lagrangian density and its parameters \citep[see, e.g.,][]{Noller:2018wyv, Huterer:2013xky}. Similarly, the analysis of the evolution of the density perturbations, at least in the linear perturbation theory \citep{DiPorto:2007ovd, BuenoSanchez:2011zz, Pace:2013pea, Kofinas:2017clq}, and of the role of the scalar field and its perturbations on structure formation are of crucial importance to test the viability of CRG.

Finally, since the scalar field drives the accelerated expansion of the Universe, we need to further investigate the connection of CRG with current dark energy models \citep{Pettorino:2005pv, Capozziello:2005mj, Frusciante:2019xia}. Specifically, we find that the parameter $w_\mathrm{DE}$ of the equation of state of the effective dark energy depends on redshift, unlike the standard cosmological model. 
When tested with measures from the upcoming Euclid mission \citep{Amendola:2016saw}, these predictions could discriminate CRG from current dark energy models. We plan to tackle all these issues in future work.

\begin{acknowledgements}
We thank Mariano Cadoni for useful discussions and an anonymous referee whose insightful comments contributed to correct and clarify some relevant aspects of the presentation of our results. An early version of this work was the Master thesis in Physics at the University of Torino of APS. We acknowledge partial support from the INFN grant InDark and the Italian Ministry of Education, University and Research (MIUR) under the {\it Departments of Excellence} grant L.232/2016. This research has made use of NASA’s Astrophysics Data System Bibliographic Services.
\end{acknowledgements}


\bibliographystyle{aa}
\bibliography{CovRG}

\newpage
\begin{appendix}

\section{Derivation of the weak-field limit}
\label{app:weak_field}

Adopting the static weak-field metric components of Eqs.~(\ref{eq:weakfield-g00})-(\ref{eq:weakfield-gij}), the Christoffel symbols and the resulting components of the Ricci tensor are
\begin{align}
\Gamma_{00}^0 &= \Gamma_{ij}^0=\Gamma_{0j}^i= 0\,, \\
\Gamma_{0i}^0 &\simeq \partial_i\Phi\,, \\
\Gamma_{00}^i &\simeq \delta^{ij}\partial_j\Phi\,, \\
\Gamma_{jk}^i &\simeq 
  -\left(\delta_k^i\partial_jU + \delta_j^i\partial_kU
      - \delta^{il}\delta_{jk}\partial_lU\right)\,, \\
R_{00} &\simeq -\nabla^2\Phi\,, \\
R_{0i} &= 0\,,\\
R_{ij} &\simeq \partial_i\partial_j(\Phi - U) - \delta_{ij}\nabla^2U \,.
\end{align}
The $\nabla_\mu\nabla_\nu\varphi$ components are
\begin{align}
& \nabla_0\nabla_0\varphi \simeq -\vec{\nabla}\Phi\cdot\vec{\nabla}\varphi, \\
& \nabla_0\nabla_i\varphi = 0, \\
& \nabla_i\nabla_j\varphi \simeq   \partial_i\partial_j\varphi
    + \partial_jU\partial_i\varphi + \partial_iU\partial_j\varphi
    - \delta_{ij}\vec{\nabla}U\cdot\vec{\nabla}\varphi \, ,
\end{align}
and
\begin{align}
&\nabla^\alpha\nabla_\alpha\varphi \simeq
  \vec{\nabla}(\Phi - U)\cdot\vec{\nabla}\varphi + \nabla^2\varphi\, .
\end{align}

For a static non-relativistic fluid, whose pressure $p$ is negligible compared to its density $\rho$, $p \ll \rho$, the equation $u^\alpha u_\alpha = -1$ implies $u_0 = (-g_{00})^{1/2}$; therefore, the components of the energy-momentum tensor $T_{\mu\nu} = \rho u_\mu u_\nu$ are
\begin{align}
T_{00} &\simeq \rho, \\
T_{0i} &= 0, \\
T_{ij} &= 0, 
\end{align}
and its trace is
\begin{align}
T_{\phantom{ij}} &\simeq -\rho.
\end{align}
Inserting these relations into Eqs.~(\ref{eq:Einstein3}) and (\ref{eq:varphimotion3}), we obtain the system of equations
\begin{align}
\vec{\nabla}\cdot\left(\varphi \vec{\nabla}\Phi\right) &\simeq 8\pi G\rho\,,
  \label{eq:wf_poisson}\\[0.2cm]
\varphi\nabla^2(\Phi - 4U)
  - \vec{\nabla}U\cdot\vec{\nabla}\varphi
  + \nabla^2\varphi - \frac{1}{\varphi}\vec{\nabla}\varphi\cdot\vec{\nabla}\varphi
  &\simeq 0\,,
  \label{eq:wf_potential}\\[0.2cm]
\vec{\nabla}(U - \Phi)\cdot\vec{\nabla}\varphi - \nabla^2\varphi
  + 2\Xi\varphi &\simeq 8\pi G\rho\,.
  \label{eq:wf_scalarfield}
\end{align}
Equation~(\ref{eq:wf_poisson}) derives from the $00$-component: it coincides with the modified Poisson equation of RG when we identify $\varphi/2$ with the permittivity $\epsilon$. Equations~(\ref{eq:wf_potential}) and (\ref{eq:wf_scalarfield}) derive from the contracted $ij$-components and the scalar field equation, respectively. Given a density distribution $\rho$, this system of equations can be solved for the three unknowns, $\Phi$, $U$, and $\varphi$. 

\subsection{The gravitational field of a spherical source immersed in a background of constant density}
\label{WF-SphSource}

For spherically-symmetric systems, the exact solution of Eq.~(\ref{eq:wf_poisson}) is
\begin{align}
\frac{d\Phi}{dr}
  &= \frac{2}{\varphi}\frac{G\int_0^r\rho\,4\pi {r^\prime}^2dr^\prime}{r^2} = \nonumber\\
  &= \frac{2}{\varphi}\frac{Gm(<r)}{r^2}\,,
\label{eq:weak_field_potential}
\end{align}
where $\rho(r)=\rho_\mathrm{s}(r)+\rho_\mathrm{bg}$, with $\rho_\mathrm{s}(r)$  being the density profile of the source in a homogeneous background with density $\rho_\mathrm{bg}$, and $m(<r)$ is the enclosed mass within radius $r$.  Newtonian gravity is recovered when $\varphi = 2$ (i.e. $\epsilon = 1$). For spherically-symmetric systems, Eqs.~(\ref{eq:wf_potential}) and (\ref{eq:wf_scalarfield}) are less straightforward to solve than Eq.~(\ref{eq:wf_poisson}). In the following, we make a few approximations to derive the generic behaviour of the gravitational field $d\Phi/dr$ and  the scalar field $\varphi$ in the two limits of small and large distances from the spherical source.

\subsubsection{The gravitational field at small distances from the spherical source}

In the Newtonian regime where $\rho_\mathrm{s}\gg\rho_\mathrm{bg}$, we may approximate $\varphi \simeq 2 - \varphi_1$, where $|\varphi_1| \ll 1$. By ignoring all the terms of the order $\mathcal{O}(\varphi_1^2)$,
$\mathcal{O}(\varphi_1)\mathcal{O}(\Phi)$, and $\mathcal{O}(\varphi_1)\mathcal{O}(U)$, Eq.~(\ref{eq:wf_scalarfield}) yields
\begin{equation}
\nabla^2\varphi_1=\frac{1}{r^2}\frac{d}{dr}\left(r^2\frac{d\varphi_1}{dr}\right)
  \simeq 8\pi G\rho \, , 
  \label{eq:wf_scalarfield_perturb}
\end{equation}
where we also ignored the terms containing $\Xi$: 
as we show in Sect.~\ref{Sec:RRGCosmology}, we can identify $\Xi$ with the cosmological constant of the standard model and, due to its small value, we can thus safely ignore these terms 
in the weak-field limit. 

Solving Eq.~(\ref{eq:wf_scalarfield_perturb}) yields the scalar field 
\begin{equation}
\varphi(r)
  \simeq 2\left[1 
    - \int_0^r\frac{Gm(<r^\prime)}{r'^2}\,dr^\prime\right]
    \,.
\label{eq:near_varphi}
\end{equation}
If the spherical source has finite size $R$ and finite mass $m$, 
Eq.~(\ref{eq:near_varphi}) yields, beyond $R$,
\begin{equation}
\varphi(r>R)
  \simeq 2 + \frac{2Gm}{r} - \frac{4\pi G\rho_\mathrm{bg} r^2}{3}\, ,
\label{eq:scalar-point-source}
\end{equation}
and Eq.~(\ref{eq:weak_field_potential}) reads as follows:
\begin{equation}
  \frac{d\Phi}{dr} \simeq \frac{2Gm}{\varphi r^2} \left(1 + \frac{4\pi \rho_\mathrm{bg}r^3}{3m}\right)\,.
\end{equation}
In the vacuum $\rho_\mathrm{bg}=0$, and the RG field reduces to the Newtonian field for $\varphi=2$, in agreement with  Eq.~(\ref{eq:wf_poisson}), and it is larger than the Newtonian field for $\varphi<2$. 

\subsubsection{The gravitational field at large distances from the spherical source}

If the mass density profile of the extended spherical source decreases with increasing $r$, at sufficiently large distances from the source, the background density $\rho_\mathrm{bg}$ approximates the mean density of the Universe, assuming that the source is isolated. We can thus treat the contribution of the source, $\rho_\mathrm{s}$, as a small perturbation to the total density field $\rho = \rho_\mathrm{bg} + \rho_\mathrm{s}$, with $\rho_\mathrm{s} \ll \rho_\mathrm{bg}$. We can similarly assume that the scalar field reaches a mean cosmic value $\varphi_\mathrm{bg}= \mathrm{const}$ and write the scalar field as $\varphi = \varphi_\mathrm{bg} + \varphi_1$, with $\varphi_1 \ll \varphi_\mathrm{bg}$.  
In the limit $\rho_\mathrm{s} \to 0$ and $\varphi_1 \to 0$,
Eqs.~(\ref{eq:wf_potential}) and (\ref{eq:wf_scalarfield}) simplify to
\begin{align}
\nabla^2U
  &\simeq \frac{1}{4}\nabla^2\Phi \simeq \frac{2\pi G \rho_\mathrm{bg}}{\varphi_\mathrm{bg}} \,, \label{eq:U_Phi} \\
\varphi_\mathrm{bg}
  &\simeq \frac{4\pi G\rho_\mathrm{bg}}{\Xi}\,.
\end{align}

We can derive how the scalar field approaches the cosmic constant value $\varphi_\mathrm{bg}$ by assuming  that Eq.~(\ref{eq:U_Phi}) already holds at distances where $\varphi$ is not yet asymptotically constant, i.e. $\nabla^2U \simeq 2\pi G \rho/\varphi$.
By adding up Eqs.~(\ref{eq:wf_potential}) and (\ref{eq:wf_scalarfield}), and using Eq.~(\ref{eq:wf_poisson}), we obtain
\begin{align}
\left(\vec{\nabla}\ln\varphi\right)^2
  + 2\vec{\nabla}\Phi\cdot\vec{\nabla}\ln\varphi
  - 2\Xi
  + \frac{8\pi G\rho}{\varphi}
  &\simeq 0\,.
\end{align}
In spherical symmetry, we can recast this equation as
\begin{equation}
\frac{d\ln\varphi}{dr}
  \simeq
  \frac{d\Phi}{dr}\left\{
  -1
  \pm \left[
    1 + \left(\frac{d\Phi}{dr}\right)^{-2}
      \left(2\Xi - \frac{8\pi G\rho}{\varphi}\right)\right]^{1/2}\right\}\, ,
\label{eq:dlnvarphidr}
\end{equation}
which yields an implicit expression for the gravitational field at large distances from the source. Hereafter, we consider the above equation with the minus sign, so that $d\varphi/dr < 0$. 

We now explore the two limits of the large and small gravitational field $d\Phi/dr$ at large distances from the source. 
In the limit $d\Phi/dr \gg (2\Xi - {8\pi G\rho}/{\varphi})^{1/2}$, Eq.~(\ref{eq:dlnvarphidr}) reduces to $d\varphi/dr \propto -2\varphi\,d\Phi/dr$, a dependence broadly approaching the result derived in the previous section for small distances from the source (Eqs.~\ref{eq:weak_field_potential} and \ref{eq:near_varphi}).

In the limit  $d\Phi/dr \ll (2\Xi - {8\pi G\rho}/{\varphi})^{1/2}$, Eq.~(\ref{eq:dlnvarphidr}) reduces to 
\begin{equation}
\frac{d\ln\varphi}{dr}\simeq - (2\Xi)^{1/2}\left(1 - \frac{4\pi G\rho}{\Xi\varphi}\right)^{1/2}\, ,
\label{eq:low-acc}
\end{equation}
which shows that $\varphi=\mathrm{const}$ when $\varphi_\mathrm{bg}=4\pi G\rho/\Xi$, and thus $\rho=\rho_\mathrm{bg}$. 

The relevant result of this analysis is that the transition between the two regimes of large and small gravitational field, $d\Phi/dr$, takes place at the acceleration scale 
\begin{equation}
\frac{d\Phi}{dr} \sim a_\Xi \equiv \left(2\Xi-\frac{8\pi G\rho}{\varphi}\right)^{1/2} \, .
\label{aXi}
\end{equation}
We discuss this result in Sect. \ref{Pointvacuumnewtonianlimit}.

At distances where $\varphi$ is not yet constant, namely $\varphi(r)=\varphi_\mathrm{bg}+\varphi_1(r)$ and  
$\rho(r)=\rho_\mathrm{bg}+\rho_\mathrm{s}(r)$, 
we can write Eq.~(\ref{eq:low-acc}) as
\begin{equation}
\frac{d\ln\varphi}{dr}\simeq - (2\Xi)^{1/2}\left(\frac{\varphi_1}{\varphi_\mathrm{bg}}
- \frac{\rho_\mathrm{s}}{\rho_\mathrm{bg}}
\right)^{1/2}\,.
\label{eq:varphi1}
\end{equation}
Solving Eq. (\ref{eq:varphi1}) for $\varphi$ requires an assumption on the form of $\rho_\mathrm{s}(r)$.  
For example,  we can assume that the density profile $\rho_\mathrm{s}(r)$ of the source drops exponentially, i.e. $\rho_\mathrm{s}(r) = \rho_\mathrm{edge}e^{-r/r_\mathrm{edge}}$, where $r_\mathrm{edge}$ is a characteristic scale of the source and $\rho_\mathrm{edge}$ is an appropriate normalization constant. In this case, $\varphi_1(r) = \varphi_\mathrm{edge}e^{-r/r_\mathrm{edge}}$, with $\varphi_\mathrm{edge} = 4\pi G\rho_\mathrm{edge}/\Xi$, is a solution of Eq.~(\ref{eq:varphi1}) if we ignore higher-order terms. We thus obtain the scalar field 
\begin{equation}
\varphi \simeq
  \frac{4\pi G}{\Xi}\left(\rho_\mathrm{bg}
  + \rho_\mathrm{edge}e^{-r/r_\mathrm{edge}}\right)\,.
\end{equation}

\section{Derivation of the modified Friedmann equations and of the scalar field equation}
\label{app:friedmann}

Based on the FLRW metric, Eq.~(\ref{eq:FLRWmetric}), the Christoffel symbols 
are\footnote{Throughout this section we only show non-zero components.}
\begin{align}
&\Gamma^0_{rr} = \frac{a \dot{a}}{1 - kr^2}\,,
  \quad \Gamma^{0}_{\theta \theta} = a \dot{a} r^2\,, \\
&\Gamma^0_{\phi \phi} = a \dot{a} r^2 \sin^2\theta\,, 
  \quad \Gamma^{r}_{0r} = \frac{\dot{a}}{a}\,,
  \quad \Gamma^{\theta}_{0\theta} = \Gamma^{\phi}_{0\phi} = \frac{\dot{a}}{a}\,, \\
&\Gamma^r_{rr} = \frac{kr}{1-kr^2}\,,
  \quad \Gamma^r_{\theta \theta} = -r (1-kr^2)\,, \\
&\Gamma^r_{\phi \phi} = -r(1-kr^2)\sin^2\theta\,,
  \quad \Gamma^{\theta}_{r\theta} = \Gamma^{\phi}_{r\phi} = \frac{1}{r}\,, \\
&\Gamma^{\theta}_{\phi \phi} = -\sin\theta \cos\theta\,,
  \quad \Gamma^{\phi}_{\theta \phi} = \frac{\cos\theta}{\sin\theta}\, .
\end{align}
The components of the Ricci tensor are
\begin{align}
R_{00} &= 3 \frac{\ddot{a}}{a}\,, \\
\label{eq:R00}
R_{rr} &= -\frac{1}{1-kr^2}\left(a \ddot{a} + 2\dot{a}^2 + 2k\right)\,, \\
\label{eq:Rrr}
R_{\theta\theta} &= -r^2\left(a \ddot{a} + 2\dot{a}^2 + 2k\right)\,, \\
R_{\phi\phi} &= -r^2\sin^2\theta\left(a \ddot{a} + 2\dot{a}^2 + 2k\right)\,,
\end{align}
and the three terms $\nabla_\mu\nabla_\nu\varphi$, $\nabla^\alpha\nabla_\alpha\varphi$, and $\nabla_\mu\varphi\nabla_\nu\varphi$ are
\begin{align}
&\nabla_0\nabla_0\varphi = \ddot{\varphi}\,,
  \quad\nabla_r\nabla_r\varphi = -\frac{1}{1 - kr^2}a\dot{a}\dot{\varphi}\,, \\
&\nabla_\theta\nabla_\theta\varphi = -r^2a\dot{a}\dot{\varphi}\,,
  \quad\nabla_\phi\nabla_\phi\varphi = -r^2\sin^2\theta a\dot{a}\dot{\varphi}\,, \\
&\nabla^\alpha\nabla_\alpha\varphi = -\ddot{\varphi} - 3\frac{\dot{a}}{a}\dot{\varphi}\,,
  \quad \nabla_0\varphi\nabla_0\varphi = \dot{\varphi}^2.
\end{align}
From the stress-energy tensor of a perfect fluid,
$T_{\mu \nu} = \left(\rho + p \right)u_{\mu} u_{\nu} + p g_{\mu \nu}$, we get
\begin{align}
&T_{00} = \rho\,, \quad T_{rr} = \frac{a}{1-kr^2}p\,, \quad T_{\theta\theta} = r^2a^2p\,,\\
&T_{\phi\phi} = r^2\sin^2\theta a^2p\,, \quad T = -(\rho - 3p)\,.
\end{align}

By combining the above results, the time-time component of the modified Einstein field equations, Eq.~(\ref{eq:Einstein3}), is
\begin{equation*}
\varphi R_{00} + \nabla_{0} \nabla_{0} \varphi -\frac{1}{\varphi} \nabla_{0} \varphi \nabla_{0} \varphi = -8\pi G T_{00}\,,
\end{equation*}
which gives the first modified Friedmann equation
\begin{equation}
\frac{\ddot{a}}{a} - \frac{1}{3}\left(\frac{\dot{\varphi}^2}{\varphi^2}-\frac{\ddot{\varphi}}{\varphi}\right) = -\frac{8\pi G}{3\varphi}\rho\,.
\label{eq:A8}
\end{equation}
From the $rr$-component of Eq.~(\ref{eq:Einstein3}),
\begin{equation}
\varphi R_{rr} + \nabla_{r} \nabla_{r} \varphi -\frac{1}{\varphi} \nabla_{r} \varphi \nabla_{r} \varphi = -8\pi G T_{rr}\,,
\label{eq:rrFriedmann}
\end{equation}
we obtain the second modified Friedmann equation
\begin{equation}
\frac{\ddot{a}}{a} + 2\frac{\dot{a}^2}{a^2} + \frac{2k}{a^2} + \frac{\dot{a}\dot{\varphi}}{a\varphi} = \frac{8\pi G}{\varphi} p\,.
\label{eq:A10}
\end{equation}
The $\theta\theta$- and $\phi\phi$-components reduce to the same expression.

The last equation is the scalar field equation
\begin{equation}
\Box \varphi - 2\Xi\varphi = 8\pi G T\,,
\label{eq:A13}
\end{equation}
which becomes
\begin{equation}
	\frac{1}{3}\frac{\ddot{\varphi}}{\varphi} + \frac{\dot{a}\dot{\varphi}}{a\varphi} + \frac{2\Xi}{3} = \frac{8\pi G}{3\varphi} \left(\rho - 3p \right)\,.
\end{equation}

\section{Derivation of the time derivative of the Hubble parameter [Eq.~(\ref{eq:dHdt})]}
\label{app:dHdt}

Adding up Eqs.~(\ref{system4a}) and (\ref{system4c}) yields
\begin{equation}
q = \Omega - 2\Omega_{\Xi} - \zeta - \frac{1}{3}\zeta^2\,.
\end{equation}
By using the definitions of $q$ and $\zeta$, Eqs.~(\ref{decparameters}) and (\ref{psi}), together with Eq.~(\ref{avarphi}), the above equation reduces to
\begin{equation}
\frac{\ddot{a}^2a^2}{2\dot{a}^4} -\frac{\ddot{a}a}{\dot{a}^2} = \frac{3}{2}\Omega - 3\Omega_{\Xi} + 1\,.
\label{diffeq}
\end{equation}
With Eqs.~(\ref{avarphi}) and (\ref{eq:density}) and the expression $\Omega(t_0) \equiv \Omega_0 = 16\pi G \rho_0/(3H_0^2\varphi_0)$, we can recast Eq.~(\ref{cosmparamOm}) as
\begin{equation}
\Omega = \frac{\Omega_0 H_0}{H}\,,
\end{equation}
and similarly (Eq.~\ref{cosmparamOl}),
\begin{equation}
\Omega_{\Xi} = \frac{\Omega_{\Xi0}H_0^2}{H^2}\,.
\end{equation}
Equation~(\ref{diffeq}) thus becomes
\begin{equation}
\frac{\ddot{a}^2}{\dot{a}^2}-\frac{2\ddot{a}}{a} - \frac{2\dot{a}^2}{a^2} = 3\Omega_0 H_0 \frac{\dot{a}}{a} - 6\Omega_{\Xi0}H_0^2\,,
\label{diffeq2}
\end{equation}
and, by adding $3\dot{a}^2/a^2$ on both sides, we obtain
\begin{equation}
\left(\frac{\ddot{a}}{\dot{a}} - \frac{\dot{a}}{a}\right)^2 = 3\frac{\dot{a}^2}{a^2} + 3\Omega_0 H_0 \frac{\dot{a}}{a} - 6\Omega_{\Xi0}H_0^2\,.
\end{equation}
The right-hand side can be rewritten as
\begin{equation}
\left(\frac{d}{dt}\ln\dot{a} - \frac{d}{dt}\ln a\right)^2
  = \left(\frac{d}{dt}\ln\frac{\dot{a}}{a}\right)^2\,,
\end{equation}
which leads to the diffential equation for $H$
\begin{equation}
\left(\frac{dH}{dt}\right)^2 = 3\left(H^4 + \Omega_0H_0H^3 - 2\Omega_{\Xi0}H_0^2H^2\right)\,.
\end{equation}

\section{Analytical derivation of the luminosity distance}
\label{sec:LumDistAnalytic}

The general definition of the luminosity distance is
\begin{equation}
D_{L} = (1+z)\int_{t}^{t_0} \frac{dt'}{a(t')}. 
\label{eq:DLRRG1}
\end{equation}
We compute its exact form by using the solution of the scale factor in Eq.~(\ref{eq:scalefactorsol}), which we rewrite here as 
\begin{eqnarray} 
 a^{\sqrt{3}} C_2^{-\sqrt{3}} &= &\left|1-2\sqrt{8\Omega_{\Xi 0}}\left\{\Omega_0 \tan \left[\sqrt{3\Omega_{\Xi 0}/2}\times \right.\right.\right. \cr
  & &  \left.\left.\left. \times\left(H_0 t+ \mathcal{C}_1 \right) \right]\mp \sqrt{\Omega_0^2 + 8\Omega_{\Xi 0}}+\sqrt{8\Omega_{\Xi 0}}\right\}^{-1} \right|\, ,
\label{5a}
\end{eqnarray}
where the upper and lower signs correspond to the CRG+ and CRG-- solutions, respectively. 

Inverting Eq.~(\ref{5a}) yields
\begin{equation}\begin{split}
&\sqrt{\frac{3}{2}\Omega_{\Xi 0}} \left(H_0 t+ C_1 \right) \equiv \arctan \left[A + \frac{B}{1-a^{\sqrt{3}} C_2^{-\sqrt{3}}} \right],
\label{ta}
\end{split}\end{equation}
where we defined
\begin{equation}
A \equiv \pm \frac{\sqrt{\Omega_0^2 + 8\Omega_{\Xi 0}}}{\Omega_0}-\frac{\sqrt{8\Omega_{\Xi 0}}}{\Omega_0}, \quad B \equiv \frac{2\sqrt{8\Omega_{\Xi 0}}}{\Omega_0}.
\label{eq:AandBdef}
\end{equation}
Differentiating Eq.~(\ref{ta}) yields
\begin{equation}\begin{split}
& dt = \frac{8}{H_0 \Omega_0}\frac{C_2^{-\sqrt{3}} a^{\sqrt{3}-1}}{\left(1-a^{\sqrt{3}} C_2^{-\sqrt{3}} \right)^2+\left(A+B-A a^{\sqrt{3}}C_2^{-\sqrt{3}} \right)^2}da.
\end{split}\end{equation} 
The luminosity distance in Eq.~(\ref{eq:DLRRG1}) then becomes
\begin{eqnarray}
D_L & = & \frac{8(1+z)}{H_0 \Omega_0}\int_0^z C_2^{-\sqrt{3}}(1+z)^{-\sqrt{3}}\times\cr
 & &\times \left\{\left[1-\left(1+z\right)^{-\sqrt{3}}C_2^{-\sqrt{3}} \right]^2 \right.+\cr
 & & + \left.\left[A+B-A\left(1+z\right)^{-\sqrt{3}}C_2^{-\sqrt{3}} \right]^2\right\}^{-1}dz.
\label{DLredshiftfirstcase}
\end{eqnarray}
If we change variable by defining
\begin{equation}\begin{split}
&y \equiv \left(1+z\right)^{-\sqrt{3}}C_2^{-\sqrt{3}}\,,
\end{split}\end{equation}
the luminosity distance becomes
\begin{equation}
D_L = \frac{8\left(1+z\right)}{\sqrt{3} C_2 H_0 \Omega_0}\int_y^{\mathcal{C}_2^{-\sqrt{3}}} \frac{y^{-1/\sqrt{3}}dy}{(1-y)^2 + (A+B-Ay)^2}\, ,
\label{eq:DL}
\end{equation}
with
\begin{eqnarray}
& & \int \frac{y^{-1/\sqrt{3}}dy}{(1-y)^2 + (A+B-Ay)^2}  =   \frac{i\sqrt{3}}{2B}y^{-1/\sqrt{3}} \times \cr
& & \times \left\{\frac{(A-i)y}{(A-i)(y-1)-B}\times \right.\cr 
& & \times\ _{2}F_{1}\left(1,1; 1+\frac{1}{\sqrt{3}}; \frac{A+B-i}{(A-i)(1-y)+B} \right) + \cr
& & -\frac{(A+i)y}{(A+i)(y-1)-B} \times\cr
& &\left. \times \ _{2}F_{1} \left(1,1; 1+\frac{1}{\sqrt{3}}; \frac{A+B+i}{(A+i)(1-y)+B} \right)\right\}.
\label{integral}
\end{eqnarray}
Equation~(\ref{eq:DL}) corresponds to two equations, according to the sign of the parameter $A$ defined in Eq.~(\ref{eq:AandBdef}). 

\section{Estimate of $\Xi$ from SNIa data}
\label{app:SNIa}

We now use the expression of the luminosity distance $D_L$ derived in the previous section and the  SNIa data from the Supernova Cosmology Project Union 2.1 Compilation \citep{Suzuki2012} 
to infer constraints on the value of $\Omega_{\Xi0}$.  We compute the quantity 
\begin{equation}
\chi^2_\nu(\Omega_0,\Omega_{\Xi0})={1\over N-2}\sum_i {[\mu_i-\mu(\Omega_0,\Omega_{\Xi0},z_i)]^2\over\sigma_{\mu i}^2}\, ,
\label{eq:chi2}
\end{equation}
where $\mu_i=m_i-M$ is the observed distance modulus, $\mu(\Omega_0,\Omega_{\Xi0},z_i)=25+5\log_{10}(D_L/{\mathrm {Mpc}})$ is the expected distance modulus at each SNIa redshift $z_i$,  $\sigma_{\mu i}$ is the uncertainty on each measured distance modulus, and the sum is over the $N=580$ SNeIa of the sample. For estimating $D_L$, we adopt $H_0 = 67.7$\,km\,s$^{-1}$\,Mpc$^{-1}$.

\begin{figure*}
\centering
\begin{subfigure}{0.45\textwidth}
\includegraphics[width=\textwidth,keepaspectratio]{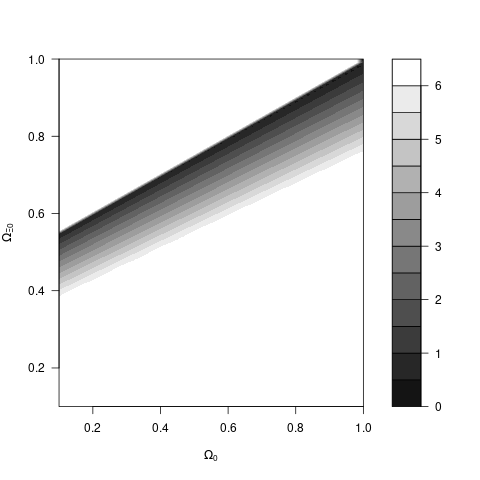}
\end{subfigure}
\hfill
\begin{subfigure}{0.45\textwidth}
\includegraphics[width=\textwidth,keepaspectratio]{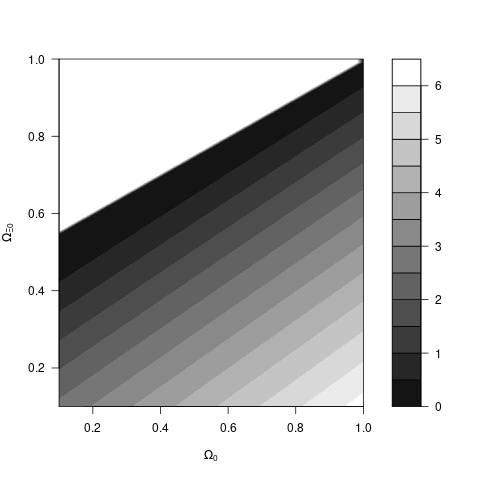}
\end{subfigure}
\caption{ Function $\Delta(\Omega_0,\Omega_{\Xi 0})=\chi^2_\nu-\chi^2_{\nu \mathrm{min}}$, the difference between $\chi^2_\nu$ (Eq. \ref{eq:chi2}) and its minimum value, for the CRG+ (left panel) and the CRG-- (right panel) solutions. Under the assumption of Gaussian random errors, the 68\%, 90\%, 95\%, and 99\% confidence regions of the two parameters are the area where $\Delta(\Omega_0,\Omega_{\Xi 0})< 2.3$, $4.6$, $6.2$, and $9.2$, respectively. The 90\% confidence regions are thus $0.43+0.44\Omega_0\lesssim \Omega_{\Xi0}\le 0.5(1+\Omega_0)$ for the CRG+ solution and $\max\{0,(-0.26+0.63\Omega_0)\}\lesssim \Omega_{\Xi0}\le 0.5(1+\Omega_0)$ for the CRG-- solution. In the two panels, there are no solutions in the white top left area. For the CRG+ solution shown in the left panel, for an easier comparison with the CRG-- solution shown in the right panel,  $\Delta(\Omega_0,\Omega_{\Xi 0})$ is not reported in the white bottom right area where $\Delta> 7$. 
}
	\label{FIG:RRG}
\end{figure*}

Figure \ref{FIG:RRG} shows that the minimum values of $\chi^2_\nu$ occur along the relation $\Omega_{\Xi0}\sim (1+\Omega_0)/2$, namely close to the upper limit of $\Omega_{\Xi0}$ (Eq. \ref{eq:cosmparambound2}). In other words, $\Omega_{\Xi0}$ is fully set by the ratio between the baryonic content of the Universe and the homogeneous scalar field $\varphi$ at the present time (Eq. \ref{cosmparamOm}). 
The result $\Omega_{\Xi0}\sim (1+\Omega_0)/2$ supports the constraints on $\Omega_{\Xi0}$ that we derived from the estimates of the dark energy parameter $w_0$  discussed in Sect. \ref{sec:EoSDE}.

\begin{figure*}
\centering
\includegraphics[width=0.45\textwidth,keepaspectratio]{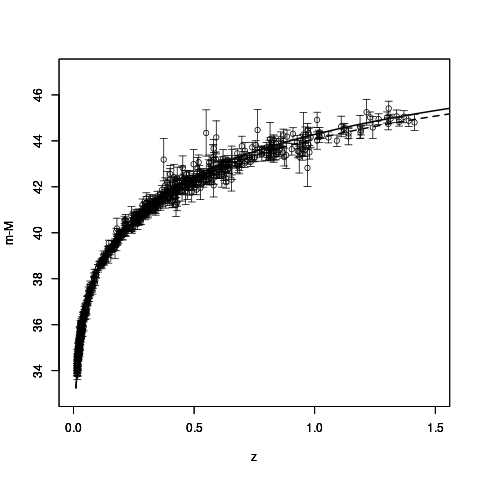}
\caption{Hubble diagram of the SNIa data (open circles with error bars) and the CRG+ (solid line) and CRG-- (dashed line) solutions.  For the two curves, we adopt $H_0 = 67.7$\,km\,s$^{-1}$\,Mpc$^{-1}$, $\Omega_0 = 0.31$, and $\Omega_{\Xi0} = 0.65$.}
	\label{FIG:HubbleDiag}
\end{figure*}

The CRG+ model yields more stringent constraints than the CRG-- model. For CRG+, the 90\% confidence limit for $\Omega_{\Xi0}$ is $0.43+0.44\Omega_0\lesssim \Omega_{\Xi0}\le 0.5(1+\Omega_0)$. 
Adopting the value $\Omega_0=0.31$ \citep{Aghanim:2018eyx},  we thus find  
$\Omega_{\Xi0}=0.650^{+0.005}_{-0.085} $ at the 90\% confidence level. 
For CRG--, the 90\% confidence limit is substantially wider: $\max\{0,(-0.26+0.63\Omega_0)\}\lesssim \Omega_{\Xi0}\le 0.5(1+\Omega_0)$. At this confidence level we find $\Omega_{\Xi0}=0.650^{+0.005}_{-0.650} $ for $\Omega_0=0.31$.
Figure \ref{FIG:HubbleDiag} compares the SNIa data in the Hubble diagram with the two CRG solutions. For the adopted values $\Omega_0=0.31$ and $\Omega_{\Xi0}=0.65$, the data are unable to distinguish between the two solutions.

\section{The equation of state of the effective dark energy}
\label{sec:EoSEffDE}

The equation of state of the effective dark energy can be calculated by inserting
Eqs.~(\ref{rhovarphi})-(\ref{pvarphi}) into
Eqs.~(\ref{rhoDEeff})-(\ref{pDEeff}).
For a dust-dominated universe, 
we can use 
Eqs.~(\ref{avarphi}) and (\ref{eq:dHdt}) to calculate $\varphi$, $\dot{H}$, and the following quantities
\begin{align}
\dot{\varphi}
  &= -\frac{H_0\varphi_0}{a^3}\left(\frac{\dot{H}}{H^2} + 3\right)\,, \\
\ddot{\varphi}
  &= \frac{H_0\varphi_0}{a^3}\left(- \frac{\ddot{H}}{H^2} + 2\frac{\dot{H}^2}{H^3}
     + 3\frac{\dot{H}}{H} + 9H\right)\,,\\
\ddot{H} &= \frac{3}{2}\left(4H^3 + 3\Omega_0 H_0 H^2 -4\Omega_{\Xi0}H_0^2 H \right)\,.
\end{align}
The combination of these results, together with the two definitions of the cosmological parameters Eqs.~(\ref{cosmparamOm})-(\ref{cosmparamOl}), yields
\begin{equation}\label{eq:wDE}
w_\mathrm{DE} =
    \frac{
    3H^2 \pm 2\sqrt{3}\left(H^4 + \Omega_0H_0H^3 - 2\Omega_{\Xi0}H_0^2H^2\right)^{1/2}}
    {-3H^2 + 3\Omega_0H_0H/2
    - \left[(Ha^3/H_0) - 1\right](3H_0^2/2a^3)\Omega_0}\,.
\end{equation}
The upper and lower signs refer to the CRG+ and the CRG-- solutions, respectively. We use this equation in Sect. \ref{sec:EoSDE}. 

\end{appendix}

\end{document}